\documentclass[a4paper,fleqn,usenatbib]{mnras}
\usepackage{newtxtext,newtxmath}
\usepackage[T1]{fontenc}
\usepackage{ae,aecompl}
\usepackage{graphicx}
\usepackage{amsmath}
\usepackage{amssymb}
\usepackage{array,multirow,makecell}
\usepackage{tabularx}
\usepackage{multirow}
\usepackage{bigints}
\usepackage{adjustbox}

\usepackage{color}
\def\gpy{{\rm ~Gpc}^{-3} {\rm ~yr}^{-1}}

\title{The connection between merging double compact objects and the Ultraluminous X-ray Sources}

\author[Mondal et al.]{
Samaresh Mondal$^{1}$\thanks{E-mail: smondal@camk.edu.pl (SM)},
Krzysztof Belczy\'nski$^{1}$, Grzegorz Wiktorowicz$^{2}$, 
\newauthor
Jean-Pierre Lasota$^{1,3}$, and Andrew~R.~King$^{3,4,5,6}$\\
$^1$Nicolaus Copernicus Astronomical Center, Polish Academy of Sciences, ul. Bartycka 18, 00-716 Warsaw, Poland\\
$^2$National Astronomical Observatories, Chinese Academy of Sciences, Beijing 100101, China\\
$^3$Institut d'Astrophysique de Paris, CNRS et Sorbonne Universite, UMR 7095, 98bis Bd Arago, 75014 Paris, France\\
$^4$Theoretical Astrophysics Group, Department of Physics and Astronomy, University of Leicester, Leicester LE1 7RH, UK\\
$^5$Astronomical Institute Anton Pannekoek, University of Amsterdam, Science Park 904, NL-1098 XH Amsterdam, the Netherlands\\
$^6$Leiden Observatory, Leiden University, Niels Bohrweg 2, NL-2333 CA Leiden, the Netherlands
}

\date{Accepted XXX. Received YYY; in original form ZZZ}
\pubyear{}

\begin{document}
\maketitle

\begin{abstract}
We explore the different formation channels of merging double compact objects (DCOs: 
BH-BH/BH-NS/NS-NS) that went through a ultraluminous X-ray phase (ULX: X-ray sources 
with apparent luminosity exceeding $10^{39}\rm{\ erg\ s^{-1}}$). 
There are many evolutionary scenarios which can naturally explain the formation of 
merging DCO systems: isolated binary evolution, dynamical evolution inside dense clusters 
and chemically homogeneous evolution of field binaries. It is not clear which scenario is 
responsible for the majority of LIGO/Virgo sources. Finding connections between ULXs and 
DCOs can potentially point to the origin of merging DCOs as more and more ULXs are 
discovered. We use the \texttt{StarTrack} population synthesis code to show how many ULXs 
will form merging DCOs in the framework of isolated binary evolution. Our merger rate 
calculation shows that in the local Universe typically $50\%$ of merging BH-BH progenitor 
binaries have evolved through a ULX phase. This indicates that ULXs can be used to study 
the origin of LIGO/Virgo sources. 
We have also estimated that the fraction of observed ULXs that will form merging DCOs in 
future varies between $5\%$ to $40\%$ depending on common envelope model and metallicity. 
\end{abstract}  


\section{Introduction}
Ultraluminous X-ray sources (ULXs) are off-nuclear point sources with apparent X-ray luminosity above 
$10^{39}\rm{\ erg\ s^{-1}}$ (see \citealt{2011NewAR..55..166F, 2017ARA&A..55..303K} for review). 
The Eddington luminosity of typical X-ray binaries (neutron star $\sim10^{38}\rm{\ erg\ s^{-1}}$ 
and a black hole of $10\ M_{\odot}\sim10^{39}\rm{\ erg\ s^{-1}}$) are below the observed luminosity 
of ULXs. ULXs were considered as potential candidates for intermediate-mass black holes ($10^2-10^{5}M_{\odot}$) 
accreting at the sub-Eddington rate \citep{1999ApJ...519...89C, Lasota et al.(2011)}, but the 
discovery of pulsating ULXs \citep{Bachetti et al.(2014), 2016ApJ...831L..14F, 2017Sci...355..817I, 
2017MNRAS.466L..48I, 2017ApJ...834...77F, 2018MNRAS.476L..45C} demonstrated that the high luminosity 
of ULXs can be achieved by super-critical accretion onto a stellar-origin compact accretor as predicted 
by \citet{King et al.(2001)}, and confirmed by \citet{2016MNRAS.458L..10K, 2017MNRAS.468L..59K, 
2019MNRAS.485.3588K} who found that the ULX luminosity results from  beamed, anisotropic emission as
suggested by \cite{King et al.(2001)}
\citep[see also][]{2019ApJ...875...53W}. Optical and near infrared observations showed that a few ULXs 
contain massive super-giant donors \citep{2007A&A...469..807L, 2011AN....332..367M, 2014Natur.514..198M, 
2015MNRAS.453.3510H, 2016MNRAS.459..771H}. Population synthesis study of field stars suggests that most 
ULXs contain $5-11\ M_{\odot}$ main sequence (MS) donors for black hole (BH) accretors and $0.9-1.5\ M_{\odot}$ 
MS donors for neutron star (NS) accretors \citep{2017ApJ...846...17W}. These donors 
indicate that many ULXs are high-mass X-ray binaries \citep{2011ApJ...741...49S,2012MNRAS.419.2095M} 
where the companion fills its Roche lobe and so transfers 
mass on a thermal timescale \citep{King et al.(2001)} and potential progenitors of close double compact objects (DCOs: BH-BH, BH-NS, NS-NS) 
\citep{2017MNRAS.472.3683F,2017A&A...604A..55M}. \cite{2018arXiv181200012K} explored a scenario of mass 
transfer from a massive donor with mass $M>15\ M_{\odot}$ onto a BH accretor leading to a ULX phase and 
eventually forming a short period BH-BH system.  

The first detection of gravitational waves (GW150914) from two merging BHs of masses around 
$\sim 30\ M_{\odot}$ was made by the advanced Laser Interferometer Gravitational-wave Observatory 
(aLIGO) \citep{2016PhRvL.116f1102A}. A total of eleven DCO mergers have been detected jointly by 
aLIGO and aVirgo during the first and second observing runs, out of which ten are BH-BH mergers 
and one is a NS-NS merger \citep{2019PhRvX...9c1040A}. \cite{2019arXiv190407214V} discovered six 
additional new  BH-BH mergers in the publicly available data from the second observing run of aLIGO/aVirgo.

There are many evolutionary scenarios which can explain the origin of BH-BH mergers: classical 
isolated binary evolution in galactic fields \citep{1993MNRAS.260..675T, 2016Natur.534..512B, 
2018MNRAS.481.1908K}, dynamical evolution inside dense star clusters \citep{2004Natur.428..724P, 
2016PhRvD..93h4029R, 2017ApJ...836L..26C, 2017MNRAS.464L..36A, 2018MNRAS.473..909B} and chemically 
homogeneous evolution of field binaries \citep{2016MNRAS.458.2634M, 2016MNRAS.460.3545D, 2016A&A...588A..50M}. 
Since we do not know yet which scenario operates for most of the BH-BH mergers, we want to find 
the potential progenitors of BH-BH mergers to constrain their origin. On the other hand, the 
connection between ULXs and merging DCOs (hereafter mDCO if their delay time is shorter than the 
Hubble age) can be used to constrain the various poorly understood physical processes in binary 
stellar evolution (efficiency of common envelope, mass transfer, natal kick distribution, etc.). 
In the classical binary evolution, most progenitors of mDCOs experience one or two mass transfer 
phases \citep{2016Natur.534..512B}. If the mass transfer rate is high enough it may lead to a 
ULX phase. We investigate a scenario in which some of the ULXs may possibly form mDCOs in the 
context of classical isolated binary evolution as proposed in earlier studies \citep{2017MNRAS.472.3683F, 
2017A&A...604A..55M, 2018arXiv181200012K}. \cite{2017MNRAS.472.3683F} did an analytical study 
assuming that all BH-BH mergers evolved through a ULX phase, which is still under debate. 
\cite{2018arXiv181200012K} explored a small range of parameter, and they only considered BH-ULXs 
with high mass donors. Our study spans a wide range of parameter space, including the most up-to-date 
prescriptions of binary stellar evolution. \cite{2012ApJ...759...52D} and \cite{2016Natur.534..512B} 
have done extensive studies of mDCOs and  predicted the current LIGO and Virgo merger rates, 
whereas \citet{2015ApJ...810...20W, 2017ApJ...846...17W, 2019ApJ...875...53W} have already drawn 
various conclusions about the population of ULXs, companion types and visibility. In this study 
we focus on the ULX formation channels that will form mDCOs at the end.

We note that the Be phenomenon \citep{1997A&A...318..443Z,1998A&A...338..505N} and formation 
of ULXs containing Be star donors are not modeled in our simulations. The formation of decretion discs
around Be stars \citep{1991MNRAS.250..432L} and the exact origin of different type of outbursts in galactic 
and extra-galactic Be stars is not yet fully understood (\cite{2001A&A...369..117N,2001A&A...369..108N}, 
but see \cite{2014ApJ...790L..34M}, who suggest that this involves Kozai--Lidov cycles in which the 
inclination of the decretion disc periodically coincides with the orbital plane, producing a massive 
outburst).  
There are at least five possible candidates of Be ULXs known at the moment; these ULXs are binary systems with 
orbital periods between 10 days to 100 days that exhibit transient phases of X-ray emission \citep{2007ApJ...663..487T,
2008MNRAS.387L..36T,2017MNRAS.471.3878T,2017A&A...605A..39T,2017ApJ...843...69W,2018MNRAS.476L..45C,
2018A&A...613A..19D,2018A&A...620L..12V}. The accretors in these systems are NSs. 
Among these system, the Be star masses are known only for two systems. NGC 300 ULX1 has a $15-25M_{\odot}$ donor 
\citep{2016MNRAS.457.1636B} and SMC X-3 has a $3.5 M_{\odot}$ donor \citep{2017MNRAS.471.3878T}. 
The donor mass in NGC 300 ULX1 is high enough that under favorable conditions, either through common envelope (CE) 
evolution or a well-placed kick, the future evolution of this system may lead to
the formation of merging NS-NS binary.

In section \ref{Simulation} we explain our simulation setup. Section \ref{accretion model} describes 
the accretion model onto compact accretors and orbital, spin parameters change due to binary interactions. 
In section \ref{beaming} we incorporate geometrical beaming in our population synthesis calculations 
in the context of ULX luminosity. We invoked two different CE models which are 
described in section \ref{CE-model}. Section \ref{result} describes our results and in section 
\ref{conclusion} we present the conclusions.

\section{Simulation}
\label{Simulation}

We used \texttt{StarTrack} \citep{2002ApJ...572..407B, 2008ApJS..174..223B}, a rapid binary 
and single star population synthesis code with major updates as described in \cite{2012ApJ...759...52D} 
and \cite{2017arXiv170607053B}. The primary (most massive) zero age main sequence (ZAMS) mass $M_{\rm a}$ 
was drawn within range $5-150\ M_{\odot}$ from three broken power-law distribution
with index $\alpha=-1.3$ for $0.08\ M_{\odot}<M_{\rm a} \le 0.5\ M_{\odot}$, $\alpha=-2.2$ 
for $0.5\ M_{\odot}<M_{\rm a}\le 1\ M_{\odot}$, and $\alpha=-2.7$ for $M_{\rm a}>1.0\ M_{\odot}$ 
\citep{1993MNRAS.262..545K}. The secondary ZAMS mass $M_{\rm b}$ ($0.5-150\ M_{\odot}$) 
was determined by the uniform distribution of binary mass ratio $q_{1}=M_{\rm b}/M_{\rm a}$ 
within range [0.1,1.0] \citep{2013A&A...550A.107S}. The orbital period ($P$) and the eccentricity 
($e$) was selected, respectively, from the distributions $f(\text{log }P/\text{d})\sim 
(\text{log }P/\rm d)^{-0.55}$ with log $P$/d in the range [0.15,5.5] and $f(e)\sim e^{-0.42}$ 
within the interval [0.0,0.9] \citep{2013A&A...550A.107S}. 

In our simulation, the rest of the 
physical assumptions are same as in the model M10 in \cite{2016A&A...594A..97B} except for the
 accretion mechanism onto a compact accretor which we explain in the next section. 
In particular, our simulation includes the rapid supernova model \citep{2012ApJ...757...91B,2012ApJ...749...91F} to estimate the mass of the final compact object after the supernova explosion. This model also 
includes the pair-instability and the pair-instability pulsation supernovae which operate for 
helium cores with masses $M_{\rm He}>60-65 M_{\odot}$ and $M_{\rm He}>40-45 M_{\odot}$, respectively
\citep[see][and references therein]{2016A&A...594A..97B}. The natal kick strength ($v_{\rm kick}$) 
during birth of a BH/NS was drawn from a Maxwellian distribution with $\sigma=265\ \rm{km\ s^{-1}}$ 
\citep{2005MNRAS.360..974H}, but decreased by the fraction of ejected mass that falls back onto the 
compact object. The final kick velocity given to a BH/NS is $v_{\rm kick,fin} = v_{\rm kick}(1-f_{\rm fb})$, 
and $f_{\rm fb}$ is the fraction of ejected mass that falls back onto the compact object.
We assumed that a BH formed via direct collapse does not receive a natal kick.

We simulated 
 $2\times10^6$ binary systems with 32 different metallicities ($Z$) from $Z=0.005Z_{\odot}$ to 
 $Z=1.5Z_{\odot}$. The exact value of $Z_{\odot}$ is not settled \citep{2017ApJ...839...55V}; we 
 adopted the value of $Z_{\odot}=0.02$. The binary fraction was chosen to be 50\% for primary 
 ZAMS mass below $10\ M_{\odot}$ and 100\%  above $10\ M_{\odot}$ \citep{2013ARA&A..51..269D, 
 2013A&A...550A.107S}. The total simulated stellar mass at each metallicity is 
$M_{\rm sim}=4.4\times10^{8}M_{\odot}$. Note that we have not used any specific 
star formation history in the context of the ULXs. In our simulation, all the stars 
are born at the same time. Our results give the total number of ULXs for a given metallicity 
that form at any time during the 10 Gyr evolution of an ensemble of stars with an initial total mass 
of $4.4\times 10^{8} M_{\odot}$.

The same simulation provides a specific number of DCOs for different metallicities. 
To calculate the cosmic merger rate density of these double compact objects as 
a function of redshift $z$, we need to use the star formation history
SFR($z$) in the Universe and the metallcity evolution as a function of redshift 
Z($z$).

SFR($z$) we adopt from \cite{2014ARA&A..52..415M},
\begin{equation}
\label{eq-sfr}
\text{SFR}(z)=0.015\frac{(1+z)^{2.7}}{1+\left(\frac{1+z}{2.9}\right)^{5.6}}\ M_{\odot}\rm{\ Mpc^{-3}\ yr^{-1}}.
\end{equation}
We calculated the merger rates from $z=0$ to 15. At each given redshift, we chose a redshift 
bin with size $\Delta z=0.1$ to calculate the comoving volume $dV_{\rm c}(z)$,
\begin{equation}
dV_{\rm c}(z)=\frac{c}{H_0}\frac{D_{\rm c}^2}{E(z)}\Delta z
\end{equation}
where $D_c$ is the comoving distance is given by,
\begin{equation}
D_{\rm c}=\frac{c}{H_0}\int_0^z\frac{dz'}{E(z')}
\end{equation}
with $E(z)=\sqrt{\Omega_{\rm M}(1+z)^3+\Omega_{\rm K}(1+z)^2+\Omega_{\Lambda}}$. $\Omega_{\rm M}$, 
$\Omega_{\rm K}$ and $\Omega_{\Lambda}$ are the usual cosmological density--parameters. The total 
stellar mass at a given redshift was determined by multiplying the SFR($z$) with $dV_{\rm c}(z)$ 
and the corresponding time interval of $\Delta z$. Then the obtained total stellar mass was used to 
normalize the simulated stellar mass. 

To include the contribution from different metallicities, 
at each redshift we used a log-normal distribution of metallicity around the average metallicity 
($Z_{\rm avg}$), with a standard deviation of $\sigma=0.5$ dex \citep{2015MNRAS.452L..36D}. The equation 
for average metallicity was taken from \cite{2014ARA&A..52..415M} with logarithmic of the 
average metallicity is increased by 0.5 dex to better fit the observational data \citep{2015MNRAS.447.2575V}
\begin{equation}
\label{eq-metallicity}
\resizebox{1.0\hsize}{!}{
$\text{log}[\text{Z}_{\rm avg}(z)]=0.5+\text{log}\bigg(\frac{y(1-R)}{\rho_{\rm b}}\bigintss_z^{20}\frac{97.8
\times10^{10}\text{SFR}(z')}{H_{0}E(z')(1+z')}dz'\bigg)$},
\end{equation}
where $y=0.019$, $R=0.27$, baryon density $\rho_{\rm b}=2.27\times10^{11}\Omega_{\rm b} h_{0}^2
\ M_{\odot}\rm{\ Mpc^{-3}}$. Throughout our study, we assumed flat cosmology with $h_{0}=0.7$,
 $\Omega_{\rm b}=0.045$, $\Omega_{\rm M}=0.3$, $\Omega_{\rm K}=0$, $\Omega_{\Lambda}=0.7$ and 
 $H_0=70.0\rm{\ km\ s^{-1}Mpc^{-1}}$.
\section{ACCRETION MODEL}
\label{accretion model}
\subsection{Roche lobe overflow (RLOF) accretion/luminosity}
\label{RLOF accretion model}
\begin{figure}
\centering
\includegraphics[width=.45\textwidth]{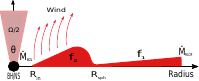}
\caption{Schematic diagram of the super-critical accretion disk around a compact accretor. 
$R_{\rm in}$ and $R_{\rm sph}$ are the inner and the spherization radius of the accretion disk. 
$f_1$ and $f_2$ parameters determine the wind mass loss rate from the outer and the inner part 
(inside $R_{\rm sph}$) of the disk. $\dot M_{\rm RLOF}$ is the mass transfer rate from the donor 
star and $\dot M_{\rm acu}$ is the mass accumulation rate onto the compact accretor. $\theta$ 
and $\Omega$ are the opening angle and the total solid angle of the emitted beam, respectively.}
\label{fig-accretion}
\end{figure}

In a close binary system when the matter is transferred from the donor star to the compact 
accretor an accretion disk is formed. We adopted the accretion disk model from 
\cite{Shakura & Sunyaev(1973)}. At low accretion rates (sub-critical) the disk does not 
produce strong outflows. At super-critical accretion rates, below the spherization radius the 
disk is dominated by radiation pressure, which leads to strong outflows. In super-critical 
accretion regime, the local disk luminosity is Eddington limited, most of the gas is blown away 
by radiation pressure and the accretion rate decreases linearly with radius (see Fig. \ref{fig-accretion}).
\par
This accretion model is used for both RLOF and wind mass accretion. First, we will discuss the 
RLOF accretion, the wind accretion is described in next section. During the Roche 
lobe overflow phase, $\dot M_{\rm RLOF}$ is the mass that has been transferred from donor star  
to the disk around compact accretor. Mass loss by the disk wind from the outer part of the disk 
down to the spherization radius ($R_{\rm sph}$) of the disk is taken care by a factor $f_1$. 
The mass accretion rate at $R_{\rm sph}$ is then
\begin{equation}
\dot M_{\rm 0,RLOF}=f_1\dot M_{\rm RLOF},
\end{equation}
but in what follows we have assumed $f_1=1$ (no wind from the outer disk). Inside the spherization 
radius ($R_{\rm sph}$), the disk is dominated by radiation pressure which leads to strong wind. 

One can calculate the spherization radius from
\begin{equation}
R_{\rm sph}=\frac{27}{4}\frac{\dot M_{\rm 0,RLOF}}{\dot M_{\rm Edd}}R_{\rm S},
\end{equation}
where $R_{\rm S}=2GM/c^2$ is the Schwarzschild radius of the accreting compact object.

The Eddington accretion rate ($\dot M_{\rm Edd}$) is given by
\begin{equation}
\dot M_{\rm Edd}\equiv \frac{L_{\rm Edd}}{0.1c^2}=4.43\times10^{-8}(1+X)^{-1}\frac{M}{M_{\odot}}
\ M_{\odot}\rm{\ yr^{-1}}
\end{equation}
where $L_{\rm Edd}=4\pi cGM/\kappa$, with $\kappa=\sigma_{\rm T}(1+X)/2m_{\rm p}$. $\sigma_{\rm T}$ 
is Thomson scattering cross-section for an electron, $m_{\rm p}$ is the mass of a proton, $G$ is 
gravitational constant and $c$ is the speed of light. The efficiency of gravitational energy release 
is $\sim$ 0.1. We take the hydrogen mass fraction in donor envelope $X$ to be 0.7 for H-rich donor 
stars and 0 for H-deficient donor stars. 
The radius of a NS ($R_{\rm NS}$) can be derived from
\begin{equation}
R_{\rm NS}=47.44-64.77\frac{M}{M_{\odot}}+39.12\left(\frac{M}{M_{\odot}}\right)^{2}-7.90\left(
\frac{M}{M_{\odot}}\right)^{3}\text{ km}
\end{equation}
where $M$ is mass of a NS. The above formula was obtained by using a polynomial fit to the data 
points of model number BSk20 from \cite{Fortin et al.(2016)}. The fit has been applied in the mass
 range from $1.39\ M_{\odot}$ to $2.17\ M_{\odot}$. We have considered the radius to be constant: 
 $R_{\rm NS}=10.37$ km for NS with masses above $2.17\ M_{\odot}$ and $R_{\rm NS}=11.77$ km for 
 NS with masses below 1.39 $M_{\odot}$.

For the case of the non-magnetized neutron star, the inner accretion disk radius, we assumed to be:
\begin{equation}
R_{\rm accNS}=R_{\rm NS},
\end{equation}
and for an accreting black hole:
\begin{equation}
R_{\rm accBH}=R_{\rm ISCO},
\end{equation}
where $R_{\rm ISCO}$ is innermost stable circular orbit radius:
\begin{equation}
R_{\rm ISCO}=\frac{GM}{c^2}\left\{ 3+Z_2 - [(3-Z_1)(3+Z_1+2Z_2)]^\frac{1}{2} \right\},
\end{equation}
where
\begin{align}
Z_2&=(3a_{\rm spin}^2+Z_1^2)^\frac{1}{2}\\
Z_1&=1+(1-a_{\rm spin}^2)^\frac{1}{3}[(1+a_{\rm spin})^\frac{1}{3}+(1-a_{\rm spin})^\frac{1}{3}]
\end{align}
where
\begin{equation}
a_{\rm spin}=\frac{Jc}{GM^2}
\end{equation}
is the BH dimensionless spin magnitude, $M$ and $J$ are respectively the mass and the spin angular 
momentum of a BH. For $a_{\rm spin}=0$, $R_{\rm ISCO}=3R_{\rm S}$. $R_{\rm ISCO}$ increases for 
retrograde motion of an orbit with respect to the BH spin, whereas in prograde motion, it comes 
closer to the horizon. We assumed the prograde rotation of the disk around the BH.

The mass accumulation rate $\dot M_{\rm acu,RLOF}$ onto the compact accretor is
\begin{equation}
\dot M_{\rm acu,RLOF}=f_2\dot M_{\rm 0,RLOF}=f_1f_2\dot M_{\rm RLOF},
\label{eq-11}
\end{equation}
where (1-$f_2$) denotes wind mass loss from the inner part of a disk (inside $R_{\rm sph}$). This 
part of the disk is assumed to be in radiation dominated regime and effectively losing mass in disk winds.
\begin{itemize}
\item If the mass transfer rate $\dot M_{\rm 0,RLOF}$ is larger than the Eddington mass accretion 
rate $\dot M_{\rm Edd}$ then 
\begin{equation}
f_{2}=\frac{R_{\rm acc}}{R_{\rm sph}}
\end{equation}
and equation \ref{eq-11} simplifies to 
\begin{equation}
\dot M_{\rm acu,RLOF}=\frac{4R_{\rm acc}}{27R_{\rm S}}\dot M_{\rm Edd}.
\end{equation}

The spherically isotropic luminosity of an accreting compact object is then given by 
\citep{Shakura & Sunyaev(1973)}
\begin{equation}
L_{\rm x,iso}=L_{\rm Edd}\bigg[1+\ln{\bigg(\frac{\dot M_{\rm 0,RLOF}}{\dot M_{\rm Edd}}\bigg)}\bigg].
\end{equation} 

\item If the  mass accretion rate $\dot M_{\rm 0,RLOF}$ is lower then the Eddington accretion rate 
then
\begin{equation}
f_{2}=1
\end{equation}
and
\begin{equation}
L_{\rm x,iso}=\eta \dot M_{\rm 0,RLOF}c^2
\end{equation}
where $\eta$ is efficiency of gravitational energy release. For NS, \citep{Shakura & Sunyaev(1973)}
\begin{equation}
\eta_{\rm NS}=\frac{GM}{c^2R_{\rm accNS}}
\end{equation}
$\eta_{\rm NS}$ varies from 17\% for 1.4 $M_{\odot}$ NS to 28\% for 2.1 $M_{\odot}$ NS. For BH,
\begin{equation}
\eta_{\rm BH}=1-E(R_{\rm ISCO})
\end{equation}
\begin{equation}
\label{eq:EISCO}
E(R)=\frac{R^2-2\frac{GM}{c^2}R + a_{\rm spin}\frac{GM}{c^2}\bigg(\frac{GM}{c^2}R\bigg)^{1/2}}
{R\bigg(R^2-3\frac{GM}{c^2}R + 2a_{\rm spin}\frac{GM}{c^2}\bigg(\frac{GM}{c^2}R\bigg)^{1/2}\bigg)^{1/2}}
\end{equation}
where $E(R=R_{\rm ISCO})$ is specific keplerian energy at ISCO radius. $\eta_{\rm BH}$ varies 
from 6\% for $a_{\rm spin}=0$ to 42\% for $a_{\rm spin}=1$.
\end{itemize}

The mass ejection rate $\dot M_{\rm eje,RLOF}$ from a disk around a compact accretor is determined by
\begin{equation}
\dot M_{\rm eje,RLOF}=\dot M_{\rm RLOF}-\dot M_{\rm acu,RLOF}.
\end{equation}

\subsection{Wind accretion/luminosity}

For the description of wind accretion we have used the \cite{Bondi & Hoyle(1944)} accretion 
mechanism. The compact accretor captures a fraction of the mass lost from the donor by stellar wind
\begin{equation}
\dot M_{\rm acc,WIND}=f_{\rm wind}\dot M_{\rm WIND},
\end{equation}
where $f_{\rm wind}$ determines the mean accretion rate into the disk around compact accretor. 
The prescription for $f_{\rm wind}$ has been taken from \cite{Hurley et al.(2002)}. Here 
$\dot M_{\rm WIND}$ is wind mass loss rate from the donor star and $\dot M_{\rm acc,WIND}$ 
is wind mass accretion rate onto the disk around the compact accretor. $f_{\rm wind}$ is given by,
\begin{equation}
\label{wind-cap}
f_{\rm wind}=\frac{1}{\sqrt{1-e^2}}\bigg(\frac{GM_{\rm acc}}{v_{\rm wind}^2}\bigg)^2\frac{
\alpha_{\rm wind}}{2a^2}\frac{1}{(1+v^2)^{3/2}}
\end{equation}
where $\alpha_{\rm wind}=1.5$, $v=v_{\rm orb}/v_{\rm wind}$ and 
\begin{equation}
v_{\rm orb}=\sqrt{\frac{G(M_{\rm acc}+M_{\rm don})}{a}}. 
\end{equation}
The wind velocity is simply assumed to be the escape velocity at the donor surface with a
 factor $\sqrt{\beta_{\rm wind}}$,
\begin{equation}
v_{\rm wind}=\sqrt{\beta_{\rm wind}\frac{2GM_{\rm don}}{R_{\rm don}}}.
\end{equation}
$\beta_{\rm wind}$ varies from 0.7 to 0.125 depending on the spectral type of the donor star. 
We  treated the rest of the problem the same way as for the RLOF accretion which translates to
\begin{equation}
\dot M_{\rm acu,WIND}=f_{\rm wind}f_1f_2\dot M_{\rm WIND}
\end{equation}
\begin{equation}
\dot M_{\rm eje,WIND}=\dot M_{\rm acc,WIND}-\dot M_{\rm acu,WIND}
\end{equation}
\begin{equation}
\dot M_{\rm 0,WIND}=f_1f_{\rm wind}\dot M_{\rm WIND}
\end{equation}
\begin{equation}
\label{wind-Lx}
\resizebox{1.0\hsize}{!}{$
L_{\rm x,iso}=
\begin{cases}
L_{\rm Edd}\bigg[1+\ln\bigg(\frac{\dot M_{\rm 0,WIND}}{\dot M_{\rm Edd}}\bigg)\bigg], & 
\text{if}\ \dot M_{\rm 0,WIND}>\dot M_{\rm Edd} \\
\eta \dot M_{\rm 0,WIND}c^2, & \text{if}\ \dot M_{\rm 0,WIND} \le \dot M_{\rm Edd} \\
\end{cases}
$}
\end{equation}
with $f_1$ and $f_2$ the same as in Section \ref{RLOF accretion model}. 

\subsubsection{Orbital parameter change}
\label{orbital}
We assumed  a spherically symmetric wind mass-loss from the donor which carries away 
the angular momentum from the binary system (Jeans-mode mass loss). This leads to orbital expansion. The 
corresponding change in orbit due to the angular momentum loss is calculated from 
\begin{equation}
a(M_{\rm acc}+M_{\rm don})=\text{constant},
\end{equation}
where only $M_{\rm don}$ changes by $\dot M_{\rm don}=(1-f_{\rm wind})\dot M_{\rm WIND}$ 
\citep{2008ApJS..174..223B}. The accumulation of mass on the compact accretor is very low 
compared to the wind mass-loss from the donor making and is not significantly affecting the 
orbital separation. In the case of super-critical accretion, the binary orbital separation 
further increases due to the wind mass loss from the inner part of the disk (inside $R_{\rm sph}$).
 We assume the matter ejected by the disk wind carries away the specific angular momentum of 
 the compact accretor. The angular momentum loss specific to the accreting compact object can 
 be obtained from
\begin{align}
\frac{dJ}{dt}&= R_{\rm com}^2\Omega_{\rm orb}\dot M_{\rm eje,RLOF/WIND}\\
R_{\rm com}&= a \frac{M_{\rm don}}{M_{\rm acc}+M_{\rm don}}\\
\Omega_{\rm orb}&= \sqrt{G(M_{\rm acc}+M_{\rm don})}a^{-1.5},
\end{align}
where $R_{\rm com}$ is the distance between the accretor and the binary's centre of mass.

\subsubsection{Compact object spin change}
The spin of the BH accretor increases due to accretion which changes the ISCO radius. The 
angular momentum $l$ and energy $E$ of the accumulated mass $M_{\rm acu}$ can be calculated 
from equation (\ref{eq:EISCO}) and from equation (3) in \cite{2008ApJ...682..474B}. Final 
mass and spin angular momentum of the BH accretor will be
\begin{align}
M_{\rm f} & =M_{\rm i}+\frac{E}{c^2}\\
J_{\rm f} & =J_{\rm i}+l
\end{align}
where the initial spin angular momentum is calculated from $J_{\rm i}=a_{\rm spin,i}M_{\rm i}^2G/c$
 and the final spin will be $a_{\rm spin,f}=J_{\rm f}c/GM_{\rm f}^2$.

\section{BEAMING MODEL}
\label{beaming}
At high mass accretion rate luminosity could be collimated through small cones then the 
observed luminosity will be much higher than $L_{\rm x,iso}$ (spherically isotropic) this
 phenomenon is called beaming \citep{King et al.(2001)}. The beaming factor $b$ has been 
 defined as $b=\Omega/4\pi$ \citep{King(2009)}. If we consider the emission through 
 two conical sections, the total solid angle of emission $\Omega=4\pi[1-\cos(\theta/2)]$, 
 here $\theta$ is the opening angle of the cone. The apparent luminosity is
\begin{center}
\begin{equation}
L_{\rm x,beam}=\frac{L_{\rm x,iso}}{b}.
\end{equation}
\end{center}
In our simulation, we identified the ULX when the apparent X-ray luminosity ($L_{\rm x,beam}$) 
of the accreting compact object exceeds $10^{39}$ erg s$^{-1}$ at some point during its lifetime. From 
comparison with observations  \cite{King(2009)} obtained for the beaming parameter $b$
\begin{equation}
b=
\begin{cases}
\frac{73}{\dot m_{0}^2}, & \dot m_{0} \ge 8.5 \\
1, & \dot m_{0} <8.5, \\
\end{cases}
\end{equation}
where, since we assume $f_1=1$, $\dot m_{0}=\dot M_{\rm 0,RLOF}/\dot M_{\rm Edd}$ is mass
 accretion rate at $R_{\rm sph}$ in Eddington accretion-rate unit.
In \citet{2017ApJ...846...17W} the beaming was assumed to saturate at very high accretion
 rates; an assumption we are not using in the present paper \citep[see][]{2019ApJ...875...53W}.

\section{Hertzsprung gap donors --- submodel A and B}
\label{CE-model}
In the scheme of close binary evolution probably the most crucial point is the CE phase. 
If the mass transfer is dynamically unstable, it will lead to a CE phase (see 
\citealt{2013A&ARv..21...59I} for review). The CE phase brings the stars closer by 
transferring the orbital energy to the envelope, which is necessary to explain the 
observed population of low mass X-ray binaries \citep[][see, however, \citet{2014bsee.confE..37W}] 
{2007A&A...469..807L} and the mDCOs \citep{2012ApJ...759...52D}. During the CE phase, 
the binary system goes through spiral--in phase, which, if the envelope is not ejected, 
will lead to a premature merger. If the donor star does not have a well developed core, 
then the orbital energy is transferred to the entire star, which makes it hard to eject 
the envelope. Stars on the MS branch do not have a clear core-envelope boundary. 
Similarly stars on the Hertzsprung gap (HG) branch lack the clear entropy difference 
related to the core-envelope structure \citep{2004ApJ...601.1058I}. We assume that a 
CE initiated by a  MS donor always result to the merger. Further we extend our 
analysis for HG donors. In submodel A, we followed the standard energy balance 
prescription of the CE for HG donors, whereas in submodel B (more conservative approach),
we assume the binary does not survive the CE initiated by HG donor. We note that systems 
such as  Cyg X-2 have avoided the CE phase despite having large mass ratio during 
the onset of mass transfer phase $q\sim2.6$ \citep{1999MNRAS.309..253K}. This type of 
system can be explained by recent study of \cite{2017MNRAS.465.2092P}, who revisited 
the stability of mass transfer and showed that at some cases the mass transfer can 
be stable even at very high mass ratio. The study by \cite{2017MNRAS.465.2092P} was 
limited to very small range of metallicities (only at $0.1Z_{\odot}$ and $Z_{\odot}$). 
We have not yet included this type of mass transfer scheme in our current study, even 
if it might explain the nature of  at least some ULXs \citep[see, e.g.][]{2019MNRAS.485.3588K}. 
In future, we will include this type of mass transfer scheme and stellar rotation in 
\texttt{StarTrack} using \texttt{MESA} model.

\section{Results}
\label{result}
\subsection{Metallicity effect on the ULX population}
\label{result-metalli}
Metallicity plays a crucial role in the binary stellar evolution. The formation number of 
ULXs can be very different at different metallicities. The numbers presented here are of 
ULXs formed out of the same stellar mass ($M_{\rm sim}=4.4\times10^8\ M_{\odot}$) at 
different metallicities. We found that ULXs can be powered by both RLOF and wind mass transfer.
Typically, RLOF ULXs are brighter than wind-fed ULXs. In general, more than $\sim 50\%$ of the 
entire RLOF ULX population have apparent luminosities larger than $10^{40}\rm{\ erg\ s^{-1}}$.
In contrast, no more than $\sim 10\%$ of all wind accreting ULXs have apparent luminosities larger 
than $10^{40}\rm{\ erg\ s^{-1}}$.

The upper panel of Fig. \ref{fig-Z_ulx} shows the number of RLOF BH- and NS-ULXs formed 
at different metallicities. For comparison we also show the total number of NS and BH binary formed. 
\subsubsection{BH-ULXs}
The number of BH-ULXs  remains almost constant at low metallicity ($0.005Z_{\odot}\le 
Z<0.2Z_{\odot}$) but decreases at higher values  (dotted blue lines). 
The mass--loss due to stellar winds plays a major role only for rather high metallicity 
which explains the relative insensitivity of the number of  ULXs formed at low metallicity values.

At higher metallicity, there are three main factors which contribute to  the decreasing 
numbers of BH-ULXs. They are: the wind mass loss, the stability properties of the 
mass-transfer and the natal kick. 

(1) The wind mass loss rate from a metal rich star is very high as compared to a 
metal-poor star \citep{2001A&A...369..574V, 2005A&A...442..587V}. Increasing wind 
mass-loss with metallicity puts the binary components further apart, which makes 
it hard to achieve the RLOF.

(2) The thermal timescale mass transfer via RLOF is allowed only when the-donor-to-accretor 
mass ratio at the onset of the RLOF is less then the critical value ($q_{\rm crit}$). If the mass ratio is 
$\geq q_{\rm crit}$, then mass transfer proceeds on dynamical timescale which leads to a CE phase. 
For rapid thermal timescale mass transfer we use a diagnostic diagram to determine $q_{\rm crit}$ 
which varies between $1.2-2.0$ depending on the type of donor \cite[Section 5.2]{2008ApJS..174..223B}. 
Stars with a radiative envelope, but with a deep convective layer are subject to delayed dynamical 
instability. \citet{1999ApJ...519L.169K} suggested that donor with radiative envelope does not lead to 
the CE phase. However, once donor convective layer is exposed it can evolve into a delayed CE phase. 
For delayed dynamical 
instability we used $q_{\rm crit}=3.0$ for H-rich donors, $q_{\rm crit}=1.7$ for He main sequence donors, 
$q_{\rm crit}=3.5$ for evolved He donors \citep{2008ApJS..174..223B}. Blue solid line in Fig. \ref{fig-Z_avgmass} 
shows the average BH mass decreases with increasing metallicity \citep{2010ApJ...714.1217B}. As metallicity 
increases the limit on the donor mass for stable mass transfer becomes narrower, which allows only a fraction 
of binary systems to go through the stable mass-transfer phase, as a result the number of RLOF BH ULXs diminishes. 

(3) The overall number of binary systems with BH accretors decreases as metallicity increases, 
which in turn lowers the number of RLOF BH ULXs (see blue dash/solid line in Fig. \ref{fig-Z_ulx}). 
The overall number of BH binary systems decreases mainly due to formation of low mass BHs. 
Low mass BHs receive natal kick during its formation, which can potentially disrupt the binary systems.

The bottom panel of Fig. \ref{fig-Z_ulx} shows the number of wind BH-ULXs, which remains 
nearly constant in all tested metallicities (dotted blue lines). This can be understood 
comparing it to the total number of binary systems formed with BH accretors. The number 
of such systems decreases with increasing metallicity, as explained in (3) above. The 
wind mass-loss rate increases with metallicity \citep{2001A&A...369..574V, 
2005A&A...442..587V}. Due to low wind mass-loss rate at low metallicity, only a fraction 
of binary systems have a mass-loss large enough to power a ULX. At high metallicity, 
although the number of companion stars that can provide the required wind mass-loss rate 
is higher, the number of binary systems with BH accretors decreases. Consequently, the number 
of wind BH-ULXs remains roughly constant throughout metallicity.

\subsubsection{NS-ULXs}
\begin{figure}
\centering
\includegraphics[width=0.5\textwidth]{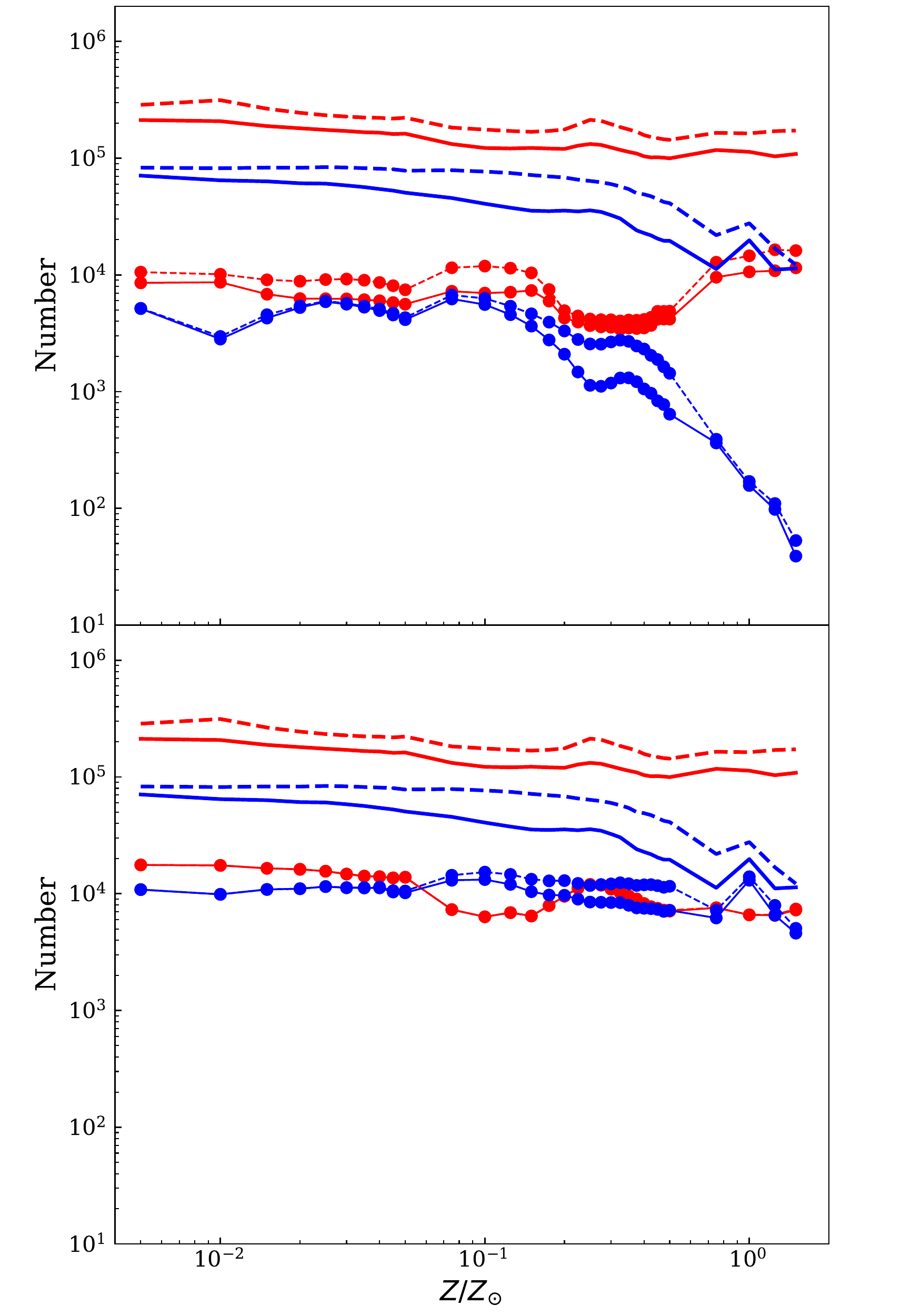}
\caption{The total number of ULXs formed in our simulation for different metallicities 
(line--connected dots). For 
comparison we also show the total number of binary systems with NS and BH accretors (lines 
with no dots). Color red corresponds to accreting NS, color blue to accreting BH. Dashed 
lines correspond to submodel A, continuous lines to submodel B. Upper panel: ULX  with 
RLOF mass transfer. Bottom Panel: ULX in wind mass transfer phase. For NS-ULXs submodels 
A and  B overlap.}
\label{fig-Z_ulx}
\end{figure}

\begin{figure}
\centering
\includegraphics[width=0.5\textwidth]{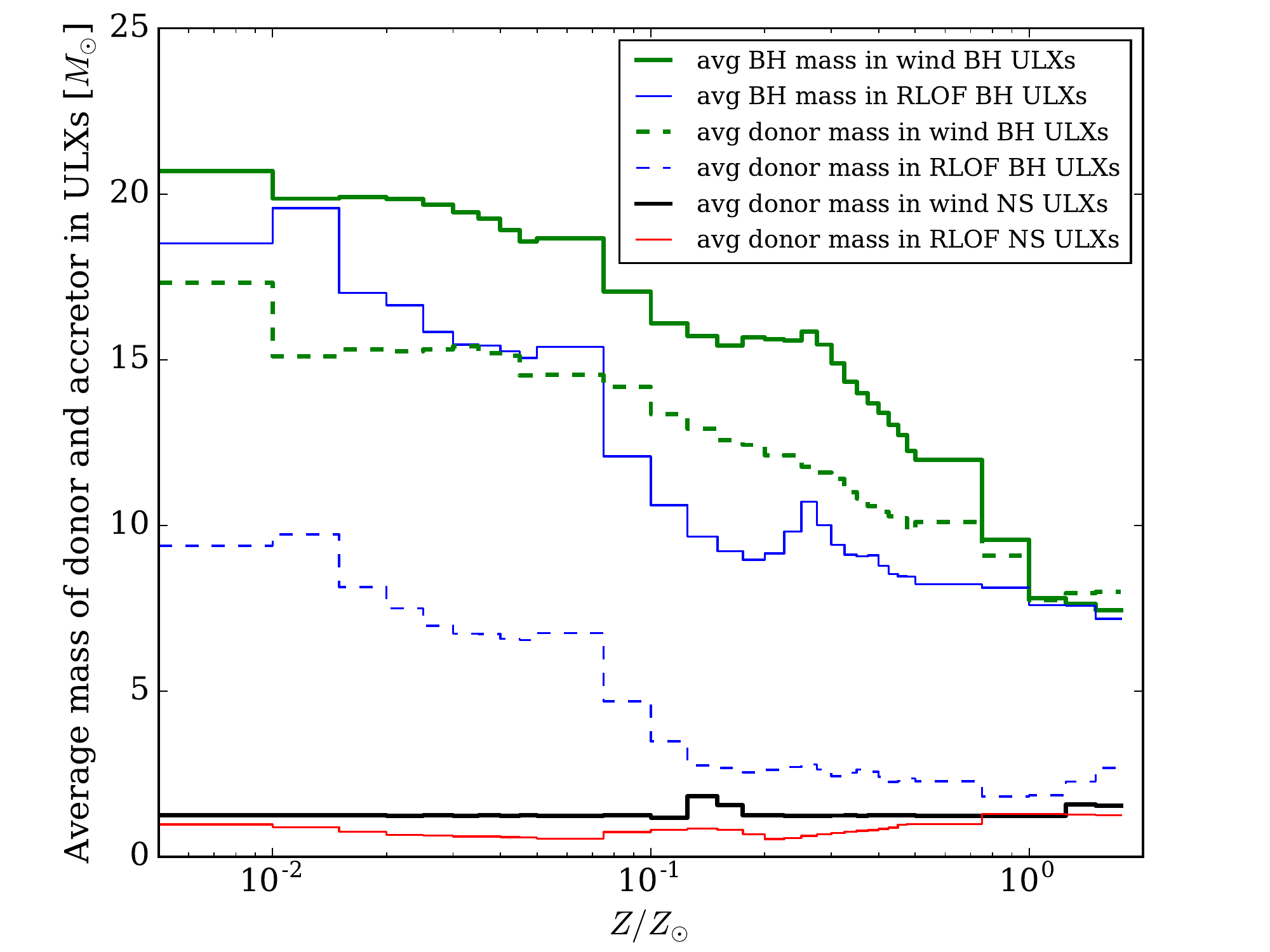}
\caption{Average masses of donors and accretors for different ULX channels in submodel 
B ( results for submodel A are similar ). The average mass of the accretor and donor 
in both type of BH-ULXs decreases as metallicity increases. The average donor mass in 
NS-ULXs remains almost constant independent of metallicity ($\sim 1.25 M_{\odot}$ for 
wind NS ULXs and $\sim 1.0 M_{\odot}$ for RLOF NS ULXs).}
\label{fig-Z_avgmass}
\end{figure}

The number of NS-ULXs does not depend much on metallicity (dashed red lines in Fig. 
\ref{fig-Z_ulx}). This is because, the donors mass in NS-ULXs are very low 
\citep{2017ApJ...846...17W, 2019ApJ...875...53W}. In our simulation, the average 
donor mass in both type of NS-ULXs are in between $1-2\ M_{\odot}$\footnote{There is 
a sub-population of high mass donor $\sim 10\ M_{\odot}$ in wind NS-ULXs with very 
small number that does not change the average mass of donor in wind NS-ULXs.} (red and 
black lines in  Fig. \ref{fig-Z_avgmass}). For low mass donors both the wind mass 
loss rates and the mass transfer rates are independent of metallicity, so their 
evolution remains nearly unaffected by metallicity.

Most NS-ULXs reach ULX luminosities through beaming of emission. For a given mass 
transfer rate, NS will always have lower opening angle of emission than BH, which 
increases the apparent luminosity of NS-ULXs \citep{2006MNRAS.366L..31K,2016MNRAS.458L..10K,2019ApJ...875...53W}. 

\subsection{Metallicity effect on the mDCOs population}
\begin{figure}
\includegraphics[width=0.5\textwidth]{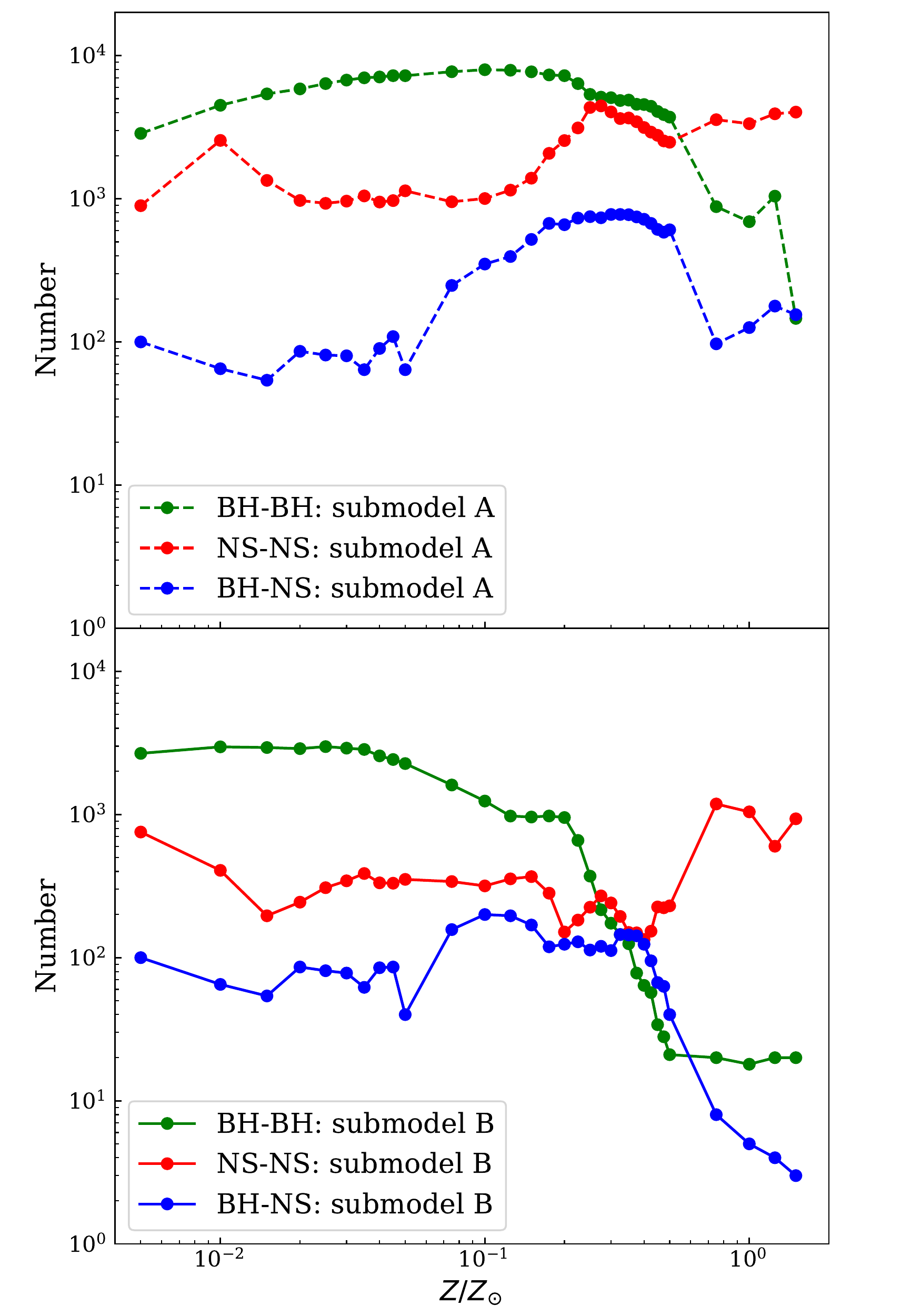}
\caption{The formation number of mDCOs at different metallicities.}
\label{fig-Z_DCO}
\end{figure}

The populations of mDCOs depend strongly on metallicity. Fig. \ref{fig-Z_DCO} 
shows the formation number of mDCOs at different metallicities. These results are well 
known from the previous studies \citep{2010ApJ...715L.138B, 2012ApJ...759...52D, 2018A&A...619A..77K}. 
The number of BH-BH formation increases with decreasing metallicity. This is mainly 
because the BH mass increases as metallicity decreases \citep{2010ApJ...714.1217B}. 
Higher mass BHs receive little to no natal kick during their formation, which leads to 
the survival of large number of binary systems. The formation efficiency of BH-NS systems 
does not increases the same way as BH-BH does with decreasing metallicity. This is because 
most of the binary systems are disrupted during the formation of NSs. The next interesting 
point to note is that the formation number (of both BH-BH and BH-NS) difference between 
submodel A and B increases with metallicity. This is because the number of BH-BH and 
BH-NS progenitors that went through CE phase with HG donors (premature merger) increases 
with metallicity \citep{2010ApJ...715L.138B}. The formation efficiency of NS-NS is less 
metallicity dependent than BH-BH and BH-NS. The natal kick strength does not change with 
metallicity for NS formation, as a NS has a very small range of mass.

\subsection{Fraction of mDCOs formed from ULXs}
\label{sec:DOCULX}
One can expect that a large fraction of mDCO evolved through an ULX phase because to 
become short period DCO these systems had to go through various phases involving very 
high mass-transfer rates \citep[see][and references therein]{2017arXiv170607053B}.

\begin{figure}
\includegraphics[width=0.5\textwidth]{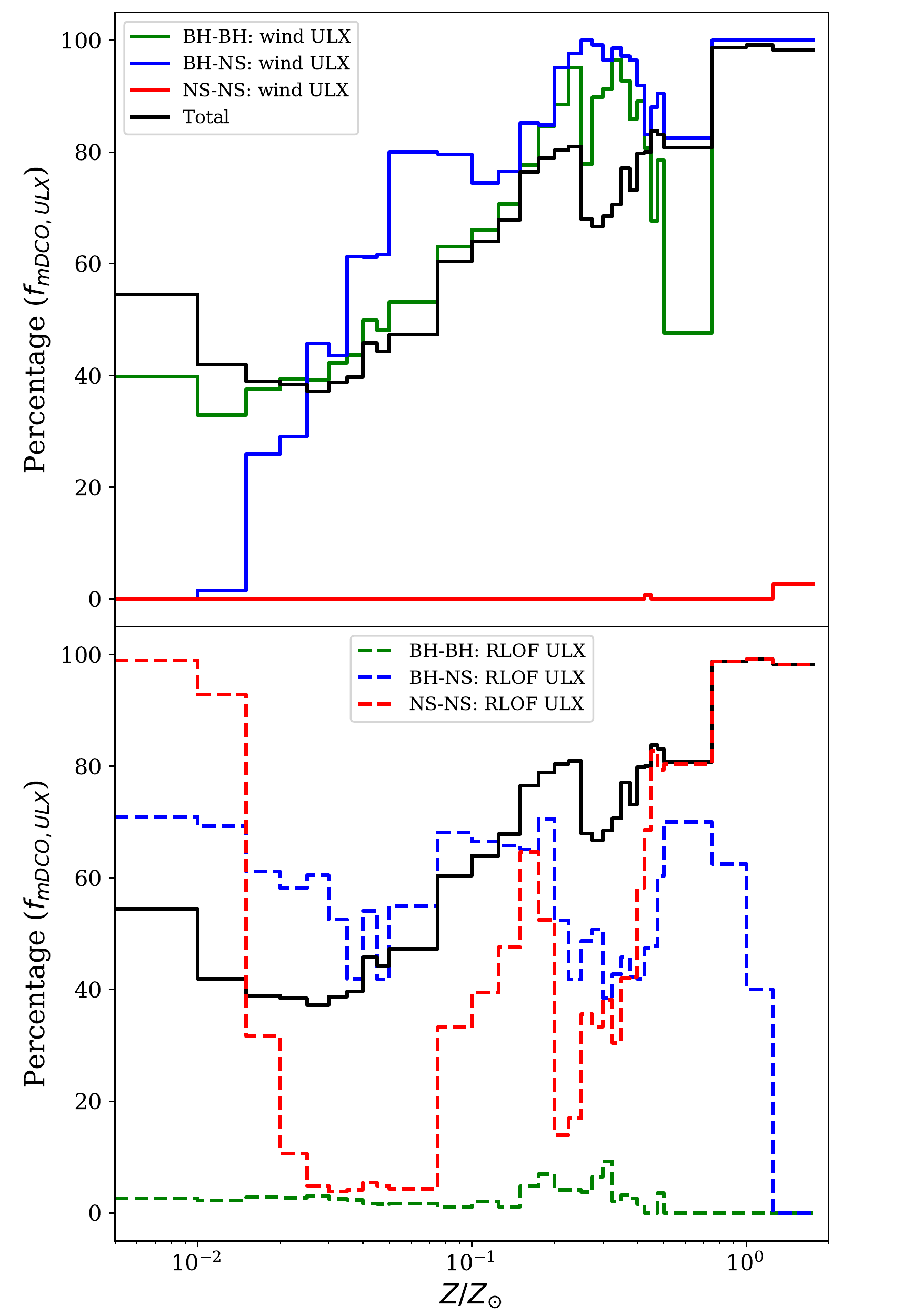}
\caption{The percentage of different mDCOs that went through different ULX phases at 
different metallicities in submodel B. Note that at some metallicities sum of wind 
and RLOF population can be higher than 100\%, this means that some ULXs went through 
both wind and RLOF mass transfer phases. The black line shows percentage of all 
mDCOs that have evolved through at least one (RLOF or wind) ULX phase.}
\label{fig-DCO_ULX}
\end{figure}

The number of mDCOs formed from ULXs channels can be very different at different 
metallicities. $f_{\rm mDCO,ULX}$ represent the percentage of mDCOs that came from 
ULX channels. For our standard model (submodel B), the values of $f_{\rm mDCO,ULX}$ 
at different metallicities are shown in Fig. \ref{fig-DCO_ULX}. The main feature here 
is that the percentage of BH-BH and BH-NS systems that went through the wind ULX phase 
increases with metallicity (upper panel of Fig. \ref{fig-DCO_ULX}). This can be 
understood using the results presented in the previous section (see section \ref{result-metalli}), 
where we showed that the population of wind BH-ULX remains nearly constant throughout 
metallicities even though the overall number of binary systems with BH accretors 
decreases at high metallicities. This indicates that as metallicity increases more 
BH binary systems have evolved through the wind ULX phase and eventually this will 
also increase the formation of BH-BH and BH-NS systems through wind ULX channel. 

In the case of the NS-NS population, almost none of the close NS-NS systems have evolved 
through the wind ULX phase. Most of the wind NS-ULXs are in wide orbits and they will 
not form merging NS-NS systems within Hubble time. 

\begin{figure}
\includegraphics[width=0.5\textwidth]{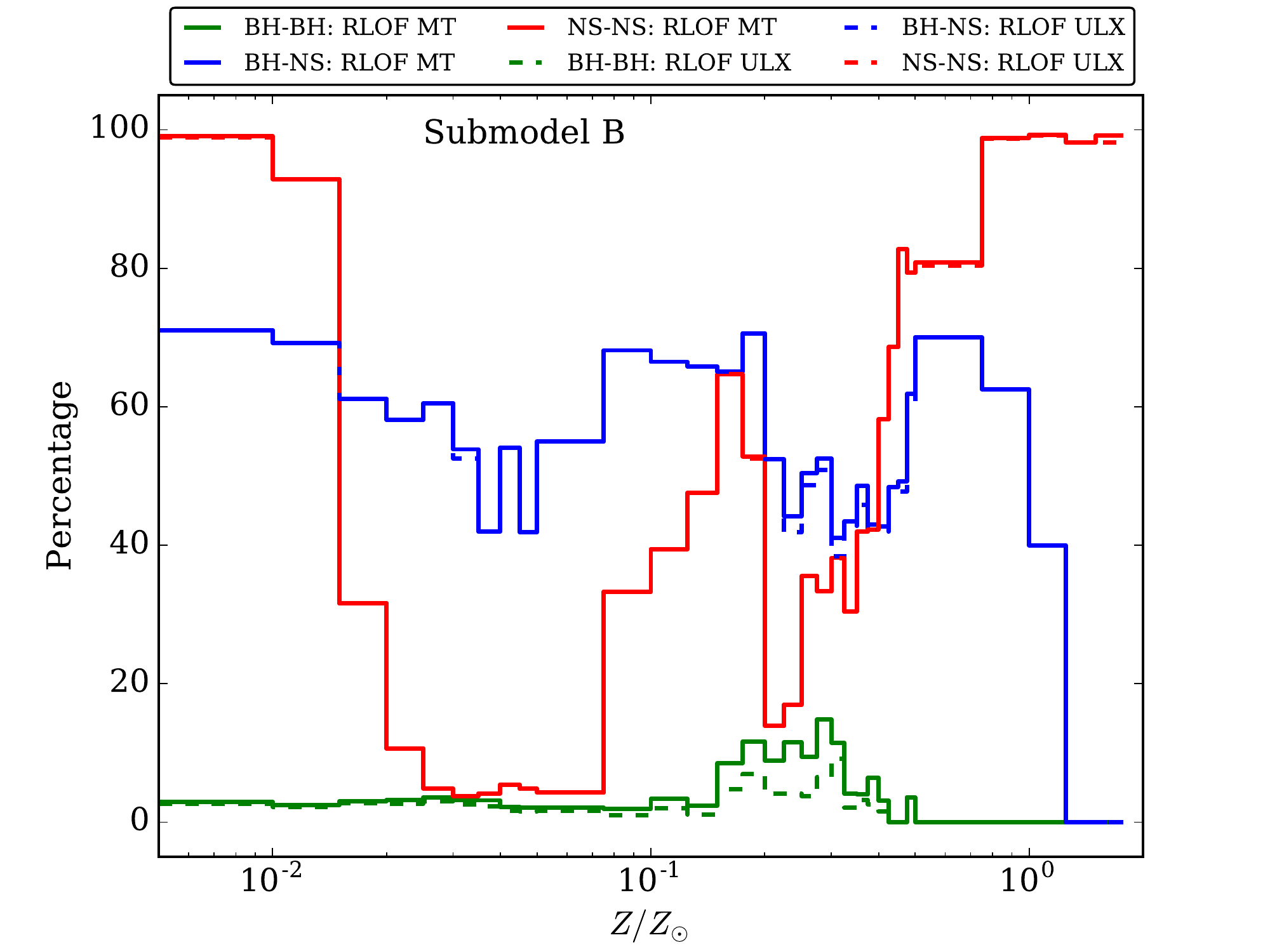}
\caption{The solid lines show the percentage of mDCOs that went through RLOF mass 
transfer phase after the first compact object formation and the dotted lines show 
the percentage of mDCOs that went through RLOF ULX phase.}
\label{fig-RLOF_ULX}
\end{figure}

The number of mDCOs that went through the RLOF ULX phase does not behave in a monotonic 
way with metallicity (bottom panel of Fig. \ref{fig-DCO_ULX}). The mDCOs that went 
through RLOF mass transfer, almost all of them achieved the ULX phase (see Fig. \ref{fig-RLOF_ULX}). 
The heavily non-monotonic behavior of $f_{\rm mDCO,ULX}$ of RLOF ULX is caused by 
various factors that change with metallicity such as the initial orbital separation 
of  DCOs\footnote{Note that the distribution of orbital separation for the whole 
population at ZAMS is same at all metallicities, but it can be very different depending 
on metallicity for the sub-population of mDCO progenitors.} \citep{2015ApJ...814...58D, 
2018A&A...619A..77K}, wind mass loss rate that changes orbital separation and radial 
expansion of the donor star \citep{2010ApJ...715L.138B}. These factors determine 
whether a given system evolves through a RLOF phase and if so, at what evolutionary 
stage. All together, these factors play a very complex role which leads to the formation 
of a non-monotonic relation between the number of RLOF systems and metallicity. 

We found that only a small percentage of merging BH-BH systems ($0-10\%$) have evolved 
through the RLOF ULX phase whereas for BH-NS and NS-NS systems the percentage, respectively, 
varies between $0-71\%$ and $4-100\%$ depending on metallicity. The small fraction of the 
ULX-descendant merging BH-BHs is due to the fact that the high mass transfer rate RLOF 
onto compact object is more restricted in case of BH-BH progenitors than for BH-NS and 
NS-NS progenitors. BH/NS can accrete at high rate (typically) either from a HG donor 
or from an evolved low-mass He-star. Massive HG stars ($\gtrsim 7\ M_{\odot}$; massive enough
to form later a NS or a BH) and low mass He stars ($\sim 2-4\ M_{\odot}$; but massive enough 
to form later NSs) are subject to significant/rapid radial expansion, leading at favorable 
binary configurations to RLOF high mass transfer rates and formation of ULXs. Massive He 
stars ($\gtrsim 4\ M_{\odot}$; that could later form BHs) do not expand significantly 
\citep{1981A&A....96..142D, 1987A&AS...69..183H, 1993RMxAA..25...79A, 1995ApJ...448..315W, 
2000MNRAS.315..543H, 2003ApJ...592..475I, 2003MNRAS.344..629D} and typically 
do not lead to high mass transfer RLOF nor to ULX phase. It follows that BH-BH progenitors 
with RLOF ULX phase are mostly restricted to HG donors, while NS-NS/BH-NS progenitors are 
allowed to have HG or low mass He star donors making it easier to generate RLOF ULX phase.

We also provide the percentage of total mDCOs that have evolved through the ULX phase 
(solid black line in Fig. \ref{fig-DCO_ULX}). The total curve nearly follows the BH-BH 
population of wind ULX at low metallicity ($Z\le0.25Z_{\odot}$). At low metallicity 
the mDCO population is dominated by BH-BH systems but as metallicity increases the 
number of BH-BH systems goes down and NS-NS becomes the major systems in the population 
of mDCOs (see Fig. \ref{fig-Z_DCO}).

\subsection{Fraction of ULXs that will form mDCOs}

We do not expect a large fraction of ULXs to become mDCO or even DCO. According 
to \citet{2017ApJ...846...17W,2019ApJ...875...53W}, ULXs have too low masses of at least 
one stellar component and/or too long orbital periods to evolve into systems that 
will be observable by LIGO/Virgo. The study by \citet{2017ApJ...846...17W,2019ApJ...875...53W} 
was limited to only RLOF ULXs, we note that, the same thing applies to wind ULXs.

Depending on the donor mass, ULXs may, or may not form mDCOs at the end of their 
evolution. $f_{\text{\tiny ULX,mDCO}}$ represents the percentage of ULXs that forms 
mDCOs out of the same simulation mass $M_{\rm sim}$.  Table \ref{Table-visDCOULX} 
shows the values of $f_{\text{\tiny ULX,mDCO}}$ for both submodels A and B. In 
submodel B, the values of $f_{\text{\tiny ULX,mDCO}}$ are very low: between 1\% to 5\% 
depending on metallicity (see also Table \ref{Table-ULXDCO}). In submodel A, 
$f_{\text{\tiny ULX,mDCO}}$ increases with metallicity, from 4\% to 15\%. As the 
different ULX populations remain nearly constant with metallicity (except for RLOF BH-ULXs), the values of $f_{\text{\tiny ULX,mDCO}}$ are simply determined by the number of mDCOs that has evolved through the ULX phase (see section \ref{sec:DOCULX}). In submodel A, 
$f_{\text{\tiny ULX,mDCO}}$ increases with metallicity because as metallicity 
increases more number of mDCOs went through the ULX phase. Whereas in submodel 
B, $f_{\text{\tiny ULX,mDCO}}$ slightly decreases with increasing metallicity simply 
because as metallicity increases more of mDCO progenitors (some of which are also ULX 
progenitors) are merged due to the CE phase initiated by an HG donor \citep{2010ApJ...715L.138B}.

Next we want to estimate what percentage of the observed ULXs will form mDCOs. Below 
we describe a model that allows to estimate the fraction of ULXs, weighted by the 
duration of ULX phase, that will eventually form mDCOs at a given metallicity. The 
probability of an ULX to be observed is directly proportional to the duration of 
ULX phase and inversely proportional to the beaming. This model utilizes only the 
beaming parameter and the lifetime of ULX phase as proxy for observability, but 
ignores the specific star formation history and the delay time between star formation 
and the onset of the ULX phase. Note that various ULXs may not only have different 
duration of high-luminosity phases, but also different delay times. Full models for 
some specific star formation history and metallicity can be easily constructed with 
our data and be used to study individual galaxies hosting ULXs. Various galaxies can 
have very complex chemical evolution and different types of star formation episodes 
(like burst type, continuous or a combination of both). Our model can only be directly 
applied to galaxies having simple properties such as a straightforward chemical 
composition and a constant star formation. $f_{\text{\tiny ULX,mDCO}}^{\text{obs}}$ 
depends both on the evolution model and the metallicity. 

We calculate $f_{\text{\tiny ULX,mDCO}}^{\text{obs}}$ (for $0.01Z_{\odot}$, 
$0.1Z_{\odot}$ and $Z_{\odot}$) as:
\begin{equation}
f_{\text{\tiny ULX,mDCO}}^{\text{obs}}=\frac{\sum\limits_{i=1}^{n}\text{dt}_\text{
\tiny ULX,mDCO}^i\times b_i}{\sum\limits_{i=1}^{n} \text{dt}_\text{\tiny ULX}^i\times b_i}\times 100\%,
\end{equation}
where the numerator represents the sum over the lifetime of ULX phase multiplied 
with the beaming parameter for ULXs that will form mDCOs at the end and the denominator 
represents the sum for all ULXs. The values of $f_{\text{\tiny ULX,mDCO}}^{\rm obs}$ are 
given Table \ref{Table-visDCOULX}. The behavior of $f_{\text{\tiny 
ULX,mDCO}}^{\rm obs}$ is much more complex than that of $f_{\text{\tiny ULX,mDCO}}$, 
as it is weighted by the duration of the ULX phase and the beaming parameter which 
are very different for different type of ULXs. RLOF ULXs tend to have longer ULX phases 
than wind ULXs. The drop of $f_{\text{\tiny ULX,mDCO}}^{\rm obs}$ at $0.1Z_{\odot}$ 
is caused by decrease in the number of mDCOs formation through RLOF ULX channel 
(shown in the bottom panel of Fig. \ref{fig-DCO_ULX}).

The duration of the ULX phase depends on the ULX accretor (BH/NS) and the ULX type 
(RLOF/wind). Table \ref{Table-timeULXphase} (see the Appendix) shows the average duration of the ULX 
phase in submodel B. The average duration of the NS-ULXs phase varies 
between $0.07-0.8$ (depending on metallicity) Myr and for BH-ULXs $0.06-0.4$ Myr. On average RLOF ULXs last $3-38$ times longer than wind ULXs.

\begin{table}
\centering
\caption{$f_{\text{\tiny ULX,mDCO}}$ represents the percentage of ULXs that has 
formed mDCOs while $f_{\text{\tiny ULX,mDCO}}^{\text{obs}}$ (weighted by the duration 
of ULX phase and beaming) represents the percentage of observed ULXs which will 
form mDCOs in future.}
\begin{adjustbox}{width=0.45\textwidth}
\setlength{\extrarowheight}{5pt}
\begin{tabular}{cccc}
\hline\hline
Model & Metallicity  & $f_{\text{\tiny ULX,mDCO}}^{\text{obs}}$ & $f_{\text{
\tiny ULX,mDCO}}$\\ \hline
\multirow{3}{*}{Submodel A} & $0.01Z_{\odot}$ & 14.0\% & 4.0\%\\
& $0.1Z_{\odot}$ & 6.9\% & 7.8\%\\
& $Z_{\odot}$ & 39.7\% & 10.8\%\\
\hline
\multirow{3}{*}{Submodel B} & $0.01Z_{\odot}$ & 14.1\% & 3.7\%\\
& $0.1Z_{\odot}$ & 4.8\% & 3.5\%\\
& $Z_{\odot}$ & 20.1\% & 3.5\%\\
\hline
\end{tabular}
\end{adjustbox}
\label{Table-visDCOULX}
\end{table}

\subsection{DCO merger rates}
\label{ULX-DCO-connec}
\begin{figure}
\centering
\includegraphics[width=0.5\textwidth]{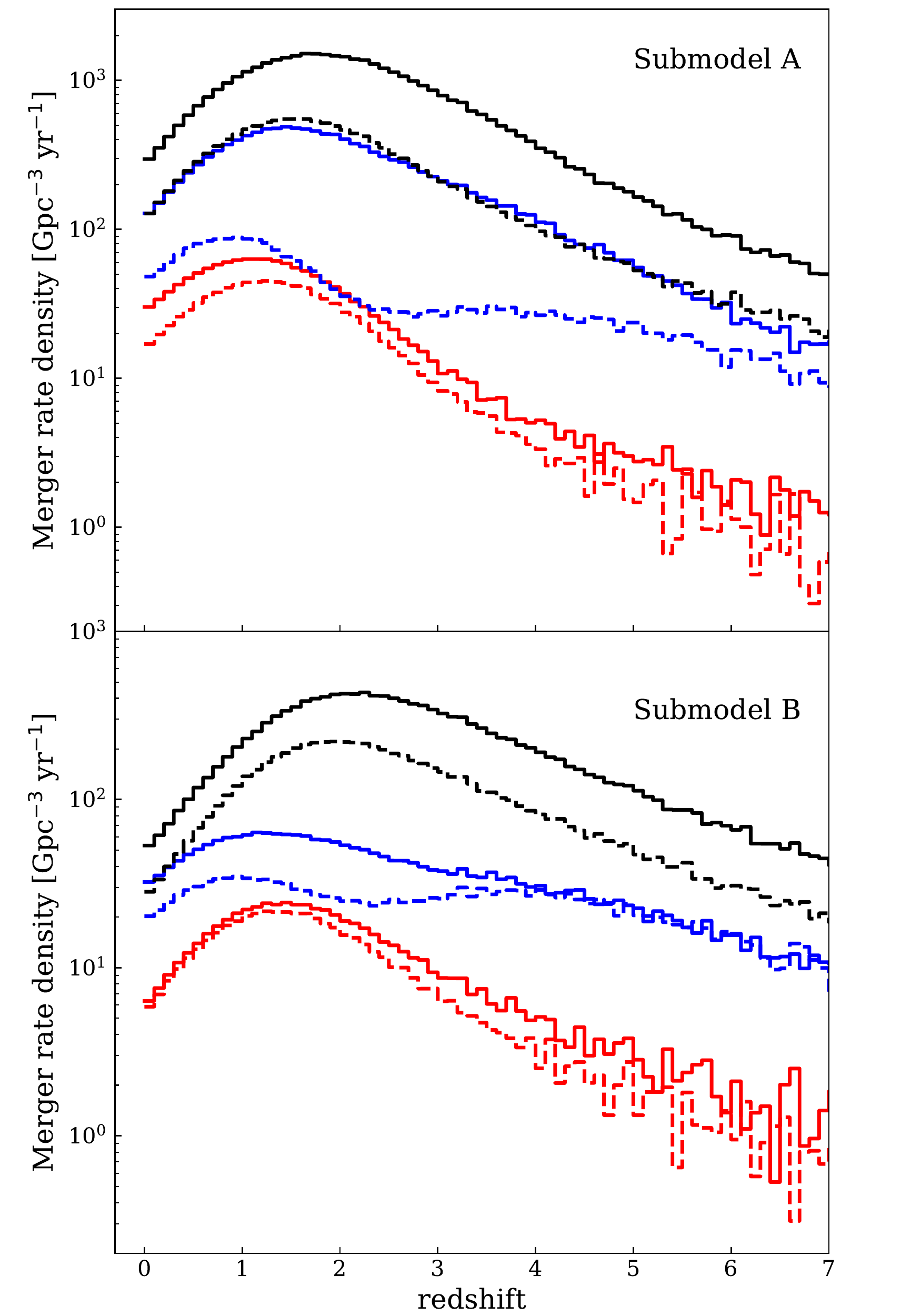}
\caption{DCO merger rate density at different redshift. Black, blue and red solid 
lines represent BH-BH, NS-NS, and BH-NS merger rate densities, respectively. Black, 
blue and red dash lines show merger rate densities of BH-BH, NS-NS and BH-NS systems 
that undergone an ULX phase in their evolution.}
\label{fig-merger-rate}
\end{figure}

We used the cosmic star formation history (Eq. \ref{eq-sfr}) and the evolution of 
average metallicity throughout cosmic time (Eq. \ref{eq-metallicity}) to calculate 
the merger rates of mDCOs ($\mathcal{R}_{\rm{mDCO}}$). Fig. \ref{fig-merger-rate} 
shows the merger rate densities at different redshift. The merger rate densities at 
the local Universe ($z=0$) are given in Table \ref{Table-Merger-rate}. Submodel A 
gives the optimistic values of merger rates, that are quite high compared to submodel B. 

Our BH-BH merger rate density ($53\gpy$ in submodel B) matches the current 
LIGO/Virgo constraint from the combined O1/O2 observational runs 
\citep[$9.7-101\gpy$;][]{2019PhRvX...9c1040A}. However, our current rates are smaller 
than the rates previously obtained with the {\tt StarTrack} code for similar evolutionary 
models \citep[e.g., model M1 submodel B in][$218\gpy$]{2016Natur.534..512B}. Note that
early (the beginning of O1) LIGO/Virgo merger rate estimate was much broader ($2-400\gpy$)
than the current O1/O2 estimate. To match the current estimate we have
changed our assumption on the IMF slope for massive stars (from $\alpha=-2.3$ to 
$\alpha=-2.7$) reducing the number of BHs in our simulations. 
A similar effect can be obtained by altering the chemical evolution model used in
calculating the merger rate densities for double compact objects (e.g., our Eq. \ref{eq-metallicity}). 
This alternative solution to matching observational estimates of the merger rates with 
{\tt StarTrack} simulations was already demonstrated by \cite{2019MNRAS.482.5012C} for the
LIGO/Virgo sources and by \cite{2019arXiv190808775O} for the Galactic populations of
double compact-object binaries. Matching the current LIGO/Virgo merger rates for
NS-NS and BH-NS mergers turns out to be more difficult than for BH-BH mergers, 
but it is achievable with various combinations of evolutionary parameters \citep[see
Fig. 25 and Fig. 26 of][]{2017arXiv170607053B}.  

Next we separately calculated the merger rate densities defined as $\mathcal{R}_{
\rm{ULX\rightarrow mDCO}}$ for systems that form mDCOs through ULX channels. Our 
merger rate calculation can be used to estimate what percentage of mDCO 
came from ULX channels. We found that in the local Universe, in submodel A, 
37\% of NS-NS, 56\% of BH-NS and 42\% of BH-BH mergers came from ULX channels, 
whereas in submodel B this percentage increases to 62\% for NS-NS, 92\% for BH-NS 
and 53\% for BH-BH. In submodel B the merger rates (both $\mathcal{R}_{\rm{mDCO}}$ 
and $\mathcal{R}_{\rm{ULX\rightarrow mDCO}}$) go down due to the merger of binary 
system during CE, initiated by HG donors. In submodel B, even though $\mathcal{R}_{
\rm{mDCO}}$ and $\mathcal{R}_{\rm{ULX\rightarrow mDCO}}$ decrease, the fraction 
$\mathcal{R}_{\rm{ULX\rightarrow mDCO}}/\mathcal{R}_{\rm{mDCO}}$ increases compared 
to submodel A (see Table \ref{Table-Merger-rate}). It indicates that lower fraction 
of ULXs went through CE phase with HG donors than the fraction of mDCOs.

\begin{table}
\centering
\caption{Estimated merger rate densities at the local Universe ($z=0$). $\mathcal{R}_{
\rm{mDCO}}$ represents the merger rate densities for mDCOs while $\mathcal{R}_{\rm{ULX
\rightarrow mDCO}}$ represents the merger rate densities for the mDCOs that are formed 
through ULX channels.}
\begin{adjustbox}{width=0.45\textwidth}
\setlength{\extrarowheight}{5pt}
\begin{tabular}{ccccc}
\hline\hline
\multirow{2}{*}{Model} & \multirow{2}{*}{DCO type} & $\mathcal{R}_{\rm{mDCO}}$ & 
$\mathcal{R}_{\rm{ULX\rightarrow mDCO}}$ & \multirow{2}{*}{Percentage}\\
&&$\rm{Gpc^{-3}\ yr^{-1}}$&$\small \rm{Gpc^{-3}\ yr^{-1}}$&\\ \hline
\multirow{3}{*}{Submodel A} & NS-NS & 128.02 & 48.25 & 37.68\%\\
& BH-NS & 30.12 & 17.02 & 56.5\%\\
& BH-BH & 296.46 & 127.25 & 42.92\%\\ \hline
\multirow{3}{*}{Submodel B} & NS-NS & 32.36 & 20.2 & 62.4\%\\
& BH-NS & 6.34 & 5.86 & 92.4\%\\
& BH-BH & 53.24 & 28.25 & 53\%\\ \hline
\end{tabular}
\end{adjustbox}
\label{Table-Merger-rate}
\end{table}

\section{Conclusions}
\label{conclusion}
We did a study of a subset of X-ray binaries -- those that went through the 
ULX phase -- and we focused on ones that form mDCOs at the end. We incorporated 
super-critical mass accretion onto a compact object and physically motivated 
beaming in our population synthesis study of large number of binary systems. 
ULX populations studied in this paper do not represent the complete sample of ULX, as 
ULXs containing Be star companions are not included in this work. The conclusions based 
on the restricted population of ULXs are listed below. 
\begin{itemize}
\item ULXs can host both NSs and BHs as accretors. The average life time of the NS-ULX 
phase varies between $0.07-0.8$ (depending on metallicity) Myr and for BH-ULX $0.06-0.4$ Myr 
(see Table \ref{Table-timeULXphase}). As NS-ULXs are more prone to be beamed 
\citep{2016MNRAS.458L..10K,2019ApJ...875...53W}, 
we obtained (weighted by beaming and life time of ULX phase) that the number of NS-ULXs would 
be $0.1-1$ (depending on metallicity) times of BH-ULXs in the observed sample of ULXs. 
Our estimate may be compared with that of \cite{2017MNRAS.470L..69M}, who found 
that in the observed sample, the number of NS-ULXs would be $\sim0.1-0.4$ times of BH-ULXs.

\item ULXs can be powered by both RLOF and wind mass transfer. In submodel B, 
on average RLOF ULXs last $3-38$ ($1-31$ times in submodel A) times longer 
than wind ULXs (see Table \ref{Table-timeULXphase}).

\item The number of RLOF BH-ULXs decreases at high metallicity while the number 
of wind BH-ULXs remains almost constant in all tested metallicities ($Z=0.005Z_{\odot}$ 
to $Z=1.5Z_{\odot}$). The number of NS-ULXs (both RLOF and wind) does not depend 
much on metallicity.

\item The average mass of donor and accretor in BH-ULXs (both RLOF and wind) decreases 
as metallicity increases. The average donor mass in RLOF BH-ULXs is 9.3 $M_{\odot}$, 
6.7 $M_{\odot}$, and 2.2 $M_{\odot}$ for $Z=0.01Z_{\odot}$, $0.1Z_{\odot}$ and $Z_{\odot}$ 
respectively. The average BH mass in RLOF BH-ULXs is 18.5 $M_{\odot}$, 15.3 $M_{\odot}$, 
and 8.2 $M_{\odot}$ for $Z=0.01Z_{\odot}$, $0.1Z_{\odot}$ and $Z_{\odot}$, respectively.

\item The average donor mass in wind and RLOF NS-ULXs is $\sim1.25\ M_{\odot}$ and 
$\sim1.0\ M_{\odot}$, respectively, almost independent of metallicity.

\item The fraction of ULXs that forms mDCOs ($f_{\text{\tiny ULX,mDCO}}$), potential 
LIGO/Virgo sources, depends both on CE outcome and metallicity. In our standard CE 
model (submodel B), the fraction is very low ($\sim3.5\%$) but in our optimistic CE 
model (submodel A) where CE events from the HG donor are allowed, the fraction is 
higher and increases with metallicity (4.0\%, 7.8\%, 10.8\% for $Z=0.01Z_{\odot}$, 
$0.1Z_{\odot}$, $Z_{\odot}$, respectively).

\item Our calculation of $f_{\text{\tiny ULX,mDCO}}^{\text{obs}}$ which is weighted 
by the duration of the ULX phase and beaming shows that $5-40\%$ (depending on CE model and 
metallicity) of the observed ULXs will form mDCOs in future.

\item From our cosmic merger rate calculation of mDCOs (see Fig. \ref{fig-merger-rate}), 
one can predict how many of the merging LIGO/Virgo sources came from ULX channels. We 
found that in the local Universe ($z=0$) the majority of the DCO mergers formed from 
isolated binaries went through a ULX phase. The numbers in two different submodel 
A/B are $37\%/62\%$ for merging NS-NS, $56\%/92\%$ for merging BH-NS and $42\%/53\%$ for merging BH-BH.
\end{itemize}

\section{ACKNOWLEDGMENTS}
We thank the anonymous referee for constructive and very useful  comments.
KB, JPL and SM acknowledge support from the Polish National Science Center (NCN) 
grants: UMO-2015/19/B/ST9/01099.
KB and SM were also partially supported by NCN Maestro grant 2018/30/A/ST9/00050. 
JPL was supported in part by a grant from the French Spatial Agency CNES. ARK thanks 
the Institut d'Astrophysique, Paris for visiting support.

\appendix
\section{SIMULATION OUTPUT}
As mentioned earlier in the paper we have simulated $2\times10^6$ binary star in 32 
different metallicity from $Z=0.005Z_{\odot}$ to $Z=1.5Z_{\odot}$. 
Table \ref{Table-timeULXphase} shows the average duration of the ULX phase. 
The detailed numerical outputs from our simulation are summarized in Table 
\ref{Table-ULXDCOfrac} and \ref{Table-ULXDCO}. Table \ref{Table-ULXDCOfrac} 
contains the formation number of different type of ULXs and DCOs. The formation 
efficiencies are also given in Table \ref{Table-ULXDCOfrac}. Table \ref{Table-ULXDCO} 
contains the most necessary informations concerning the connection between ULX and DCO. 
The percentage of ULXs that ends up forming DCOs and the percentage of DCOs that 
came from ULX channels both numbers are given in Table \ref{Table-ULXDCO}.

\begin{table*}
\setlength\extrarowheight{5pt}
\centering
\caption{}{The average lifetime of the ULX phase for different type of ULXs in submodel B. 
$\tau$ is the average lifetime of the ULX phase in Myr. RLOF and wind ULX represent 
the mass transfer mode in ULX. The subscript of $\tau$ denotes the accretor type in ULXs.}
\begin{center}
\scalebox{1.0}{
\begin{tabular}{|c|c|c|c|c|c|c|c|c|}
\hline
\multirow{2}{*}{Z} & \multicolumn{3}{c|}{RLOF ULX} & \multicolumn{3}{c|}{wind ULX} & 
\multirow{2}{*}{$\tau_{\rm NS(RLOF+wind)}$} & \multirow{2}{*}{$\tau_{\rm BH(RLOF+wind)}$}
\\ \cline{2-7}
& $\tau_{\rm NS}$ & $\tau_{\rm BH}$ & $\tau_{\rm NS+BH}$ & $\tau_{\rm NS}$ & $\tau_{
\rm BH}$ & $\tau_{\rm NS+BH}$&&\\ \hline

0.005$Z_{\odot}$ & 0.210 & 1.046 & 0.524 & 0.016 & 0.104 & 0.049 & 0.079 & 0.407 \\ \hline
0.01$Z_{\odot}$ & 0.174 & 0.909 & 0.354 & 0.016 & 0.128 & 0.056 & 0.068 & 0.301 \\ \hline
0.015$Z_{\odot}$ & 0.641 & 1.247 & 0.874 & 0.017 & 0.104 & 0.052 & 0.200 & 0.426 \\ \hline
0.02$Z_{\odot}$ & 1.108 & 0.846 & 0.988 & 0.018 & 0.095 & 0.049 & 0.322 & 0.337 \\ \hline
0.025$Z_{\odot}$ & 1.433 & 0.834 & 1.143 & 0.019 & 0.091 & 0.050 & 0.424 & 0.343 \\ \hline
0.03$Z_{\odot}$ & 1.532 & 0.850 & 1.209 & 0.020 & 0.089 & 0.050 & 0.469 & 0.343 \\ \hline
0.035$Z_{\odot}$ & 1.918 & 0.887 & 1.441 & 0.021 & 0.090 & 0.051 & 0.594 & 0.345 \\ \hline
0.04$Z_{\odot}$ & 1.976 & 1.091 & 1.576 & 0.020 & 0.092 & 0.052 & 0.608 & 0.398 \\ \hline
0.045$Z_{\odot}$ & 2.645 & 1.078 & 1.957 & 0.020 & 0.094 & 0.052 & 0.802 & 0.394 \\ \hline
0.05$Z_{\odot}$ & 2.018 & 1.337 & 1.729 & 0.020 & 0.096 & 0.052 & 0.596 & 0.454 \\ \hline
0.075$Z_{\odot}$ & 0.876 & 1.056 & 0.959 & 0.031 & 0.062 & 0.051 & 0.451 & 0.383 \\ \hline
0.1$Z_{\odot}$ & 0.788 & 1.214 & 0.977 & 0.030 & 0.065 & 0.053 & 0.427 & 0.405 \\ \hline
0.125$Z_{\odot}$ & 0.655 & 1.405 & 0.948 & 0.029 & 0.072 & 0.056 & 0.347 & 0.439 \\ \hline
0.15$Z_{\odot}$ & 0.692 & 1.400 & 0.926 & 0.032 & 0.093 & 0.070 & 0.384 & 0.432 \\ \hline
0.175$Z_{\odot}$ & 1.277 & 1.726 & 1.419 & 0.029 & 0.115 & 0.076 & 0.565 & 0.472 \\ \hline
0.2$Z_{\odot}$ & 1.186 & 1.692 & 1.352 & 0.024 & 0.131 & 0.078 & 0.384 & 0.408 \\ \hline
0.225$Z_{\odot}$ & 0.752 & 1.836 & 1.047 & 0.020 & 0.126 & 0.067 & 0.211 & 0.367 \\ \hline
0.25$Z_{\odot}$ & 0.535 & 2.356 & 0.964 & 0.017 & 0.126 & 0.062 & 0.139 & 0.388 \\ \hline
0.275$Z_{\odot}$ & 0.612 & 2.367 & 1.027 & 0.018 & 0.117 & 0.059 & 0.156 & 0.379 \\ \hline
0.3$Z_{\odot}$ & 0.378 & 1.966 & 0.774 & 0.019 & 0.105 & 0.056 & 0.107 & 0.335 \\ \hline
0.325$Z_{\odot}$ & 0.537 & 1.751 & 0.873 & 0.021 & 0.104 & 0.058 & 0.150 & 0.326 \\ \hline
0.35$Z_{\odot}$ & 0.698 & 1.486 & 0.913 & 0.022 & 0.099 & 0.057 & 0.205 & 0.296 \\ \hline
0.375$Z_{\odot}$ & 0.665 & 1.560 & 0.897 & 0.024 & 0.096 & 0.057 & 0.203 & 0.299 \\ \hline
0.4$Z_{\odot}$ & 0.734 & 1.526 & 0.917 & 0.025 & 0.089 & 0.056 & 0.239 & 0.266 \\ \hline
0.425$Z_{\odot}$ & 0.710 & 1.349 & 0.843 & 0.027 & 0.082 & 0.054 & 0.250 & 0.227 \\ \hline
0.45$Z_{\odot}$ & 0.645 & 0.990 & 0.703 & 0.028 & 0.077 & 0.052 & 0.251 & 0.170 \\ \hline
0.475$Z_{\odot}$ & 0.534 & 0.801 & 0.576 & 0.028 & 0.073 & 0.050 & 0.215 & 0.145 \\ \hline
0.5$Z_{\odot}$ & 0.608 & 0.597 & 0.607 & 0.028 & 0.069 & 0.049 & 0.243 & 0.112 \\ \hline
0.75$Z_{\odot}$ & 0.150 & 1.689 & 0.206 & 0.028 & 0.060 & 0.042 & 0.096 & 0.150 \\ \hline
$Z_{\odot}$ & 0.150 & 1.387 & 0.168 & 0.030 & 0.055 & 0.046 & 0.104 & 0.071 \\ \hline
1.25$Z_{\odot}$ & 0.124 & 0.505 & 0.127 & 0.030 & 0.064 & 0.047 & 0.089 & 0.070 \\ \hline
1.5$Z_{\odot}$ & 0.123 & 0.337 & 0.124 & 0.027 & 0.058 & 0.039 & 0.086 & 0.060 \\ \hline
\end{tabular}}
\end{center}
\label{Table-timeULXphase}
\end{table*}

\begin{table*}
\setlength\extrarowheight{5pt}
\centering
\caption{}{The number of different systems formed from simulation of $2\times10^6$ binary 
stars at each metallicity in submodel B. The corresponding simulation mass is 
$M_{\rm sim}=4.4\times10^8M_{\odot}$ . ULX$^R$ and ULX$^W$ represent the number of ULX systems 
formed during RLOF and wind mass transfer episodes, respectively. NS-NS, BH-NS and BH-BH 
represent the number of mDCOs.}
\begin{center}
\scalebox{0.85}{
\begin{tabular}{|c|c|c|c|c|c|c|c|c|c|c|c|c|}

\hline

Z & NS-ULX$^R$ & BH-ULX$^R$ & $\frac{\text{ULX$^R$}}{M_{\rm sim}}$ & NS-ULX$^W$ & BH-ULX$^W$ 
& $\frac{\text{ULX$^W$}}{M_{\rm sim}}$ & NS-NS & $\frac{\text{NS-NS}}{M_{\rm sim}}$ & BH-NS & 
$\frac{\text{BH-NS}}{M_{\rm sim}}$ & BH-BH & $\frac{\text{BH-BH}}{M_{\rm sim}}$\\ \hline

0.005$Z_{\odot}$ & 8545 & 5141 & 3.1e-05 & 17638 & 10846 & 6.4e-05 & 754 & 1.7e-06 & 100 & 2.2e-07 & 2672 & 6e-06\\ \hline
0.01$Z_{\odot}$ & 8646 & 2818 & 2.6e-05 & 17483 & 9878 & 6.2e-05 & 407 & 9.1e-07 & 65 & 1.5e-07 & 2967 & 6.7e-06\\ \hline
0.015$Z_{\odot}$ & 6830 & 4267 & 2.5e-05 & 16499 & 10878 & 6.2e-05 & 196 & 4.4e-07 & 54 & 1.2e-07 & 2935 & 6.6e-06\\ \hline
0.02$Z_{\odot}$ & 6266 & 5248 & 2.6e-05 & 16186 & 11037 & 6.1e-05 & 244 & 5.5e-07 & 86 & 1.9e-07 & 2884 & 6.5e-06\\ \hline
0.025$Z_{\odot}$ & 6248 & 5886 & 2.7e-05 & 15590 & 11507 & 6.1e-05 & 308 & 6.9e-07 & 81 & 1.8e-07 & 2979 & 6.7e-06\\ \hline
0.03$Z_{\odot}$ & 6226 & 5612 & 2.7e-05 & 14739 & 11236 & 5.8e-05 & 344 & 7.7e-07 & 78 & 1.8e-07 & 2907 & 6.5e-06\\ \hline
0.035$Z_{\odot}$ & 6130 & 5288 & 2.6e-05 & 14159 & 11198 & 5.7e-05 & 387 & 8.7e-07 & 62 & 1.4e-07 & 2846 & 6.4e-06\\ \hline
0.04$Z_{\odot}$ & 6006 & 4947 & 2.5e-05 & 13989 & 11216 & 5.7e-05 & 333 & 7.5e-07 & 85 & 1.9e-07 & 2567 & 5.8e-06\\ \hline
0.045$Z_{\odot}$ & 5796 & 4535 & 2.3e-05 & 13664 & 10337 & 5.4e-05 & 331 & 7.4e-07 & 86 & 1.9e-07 & 2421 & 5.4e-06\\ \hline
0.05$Z_{\odot}$ & 5610 & 4132 & 2.2e-05 & 13849 & 10177 & 5.4e-05 & 352 & 7.9e-07 & 40 & 9e-08 & 2268 & 5.1e-06\\ \hline
0.075$Z_{\odot}$ & 7240 & 6220 & 3e-05 & 7308 & 13036 & 4.6e-05 & 340 & 7.6e-07 & 157 & 3.5e-07 & 1607 & 3.6e-06\\ \hline
0.1$Z_{\odot}$ & 6982 & 5569 & 2.8e-05 & 6343 & 13215 & 4.4e-05 & 317 & 7.1e-07 & 200 & 4.5e-07 & 1241 & 2.8e-06\\ \hline
0.125$Z_{\odot}$ & 7108 & 4562 & 2.6e-05 & 6886 & 12021 & 4.2e-05 & 355 & 8e-07 & 196 & 4.4e-07 & 975 & 2.2e-06\\ \hline
0.15$Z_{\odot}$ & 7350 & 3641 & 2.5e-05 & 6442 & 10408 & 3.8e-05 & 368 & 8.3e-07 & 169 & 3.8e-07 & 960 & 2.2e-06\\ \hline
0.175$Z_{\odot}$ & 5961 & 2770 & 2e-05 & 7917 & 9745 & 4e-05 & 282 & 6.3e-07 & 119 & 2.7e-07 & 976 & 2.2e-06\\ \hline
0.2$Z_{\odot}$ & 4270 & 2092 & 1.4e-05 & 9514 & 9701 & 4.3e-05 & 151 & 3.4e-07 & 124 & 2.8e-07 & 952 & 2.1e-06\\ \hline
0.225$Z_{\odot}$ & 3944 & 1473 & 1.2e-05 & 11146 & 8989 & 4.5e-05 & 183 & 4.1e-07 & 129 & 2.9e-07 & 659 & 1.5e-06\\ \hline
0.25$Z_{\odot}$ & 3678 & 1132 & 1.1e-05 & 11995 & 8483 & 4.6e-05 & 225 & 5.1e-07 & 113 & 2.5e-07 & 371 & 8.3e-07\\ \hline
0.275$Z_{\odot}$ & 3587 & 1111 & 1.1e-05 & 11834 & 8444 & 4.6e-05 & 270 & 6.1e-07 & 120 & 2.7e-07 & 216 & 4.9e-07\\ \hline
0.3$Z_{\odot}$ & 3561 & 1184 & 1.1e-05 & 10938 & 8404 & 4.3e-05 & 241 & 5.4e-07 & 112 & 2.5e-07 & 174 & 3.9e-07\\ \hline
0.325$Z_{\odot}$ & 3416 & 1308 & 1.1e-05 & 10182 & 8386 & 4.2e-05 & 194 & 4.4e-07 & 145 & 3.3e-07 & 145 & 3.3e-07\\ \hline
0.35$Z_{\odot}$ & 3507 & 1312 & 1.1e-05 & 9469 & 7966 & 3.9e-05 & 150 & 3.4e-07 & 144 & 3.2e-07 & 125 & 2.8e-07\\ \hline
0.375$Z_{\odot}$ & 3472 & 1216 & 1.1e-05 & 8931 & 7559 & 3.7e-05 & 149 & 3.3e-07 & 142 & 3.2e-07 & 78 & 1.8e-07\\ \hline
0.4$Z_{\odot}$ & 3511 & 1055 & 1e-05 & 8139 & 7500 & 3.5e-05 & 134 & 3e-07 & 124 & 2.8e-07 & 64 & 1.4e-07\\ \hline
0.425$Z_{\odot}$ & 3690 & 968 & 1e-05 & 7623 & 7451 & 3.4e-05 & 153 & 3.4e-07 & 95 & 2.1e-07 & 57 & 1.3e-07\\ \hline
0.45$Z_{\odot}$ & 4169 & 834 & 1.1e-05 & 7372 & 7338 & 3.3e-05 & 226 & 5.1e-07 & 67 & 1.5e-07 & 34 & 7.6e-08\\ \hline
0.475$Z_{\odot}$ & 4184 & 775 & 1.1e-05 & 7131 & 7060 & 3.2e-05 & 223 & 5e-07 & 63 & 1.4e-07 & 28 & 6.3e-08\\ \hline
0.5$Z_{\odot}$ & 4177 & 641 & 1.1e-05 & 7092 & 7199 & 3.2e-05 & 230 & 5.2e-07 & 40 & 9e-08 & 21 & 4.7e-08\\ \hline
0.75$Z_{\odot}$ & 9545 & 363 & 2.2e-05 & 7573 & 6196 & 3.1e-05 & 1184 & 2.7e-06 & 8 & 1.8e-08 & 20 & 4.5e-08\\ \hline
$Z_{\odot}$ & 10617 & 157 & 2.4e-05 & 6593 & 13048 & 4.4e-05 & 1043 & 2.3e-06 & 5 & 1.1e-08 & 18 & 4e-08\\ \hline
1.25$Z_{\odot}$ & 10850 & 98 & 2.5e-05 & 6533 & 6529 & 2.9e-05 & 600 & 1.3e-06 & 4 & 9e-09 & 20 & 4.5e-08\\ \hline
1.5$Z_{\odot}$ & 11496 & 39 & 2.6e-05 & 7245 & 4591 & 2.7e-05 & 933 & 2.1e-06 & 3 & 6.7e-09 & 20 & 4.5e-08\\ \hline

\end{tabular}}
\end{center}
\label{Table-ULXDCOfrac}
\end{table*}

\begin{table*}
\setlength\extrarowheight{5pt}
\centering
\caption{}{NS-NS, NS-BH and BH-BH represent the number of mDCOs that went through an ULX 
phase. ULX$^R$ and ULX$^W$ represent RLOF and wind ULX phases, respectively. 4th column 
shows the number of systems that went through both ULX$^R$ and ULX$^W$ phases. 
$f_{\text{\tiny ULX,mDCO}}$ shows the percentage of ULXs that forms mDCOs and 
$f_{\text{\tiny mDCO,ULX}}$ shows what percentage of mDCOs came from ULX channels. 
This table has been given for submodel B.}
\begin{center}
\scalebox{1.0}{
\begin{tabular}{|c|c|c|c|c|c|c|c|c|c|c|c|}

\hline
\multirow{2}{*}{Z} & \multicolumn{3}{c|}{ULX$^R$} & \multicolumn{3}{c|}{ULX$^W$} & 
\multicolumn{3}{c|}{ULX$^R$ and ULX$^W$} & \multirow{2}{*}{$f_{\text{\tiny ULX,mDCO}}$} 
& \multirow{2}{*}{$f_{\text{\tiny mDCO,ULX}}$}\\
\cline{2-10}
& NS-NS & BH-NS & BH-BH & NS-NS & BH-NS & BH-BH & NS-NS & BH-NS & BH-BH & &\\ \hline

0.005$Z_{\odot}$ & 746 & 71 & 70 & 0 & 0 & 1063 & 0 & 0 & 29 & 4.6\% & 54.8\%\\ \hline
0.01$Z_{\odot}$ & 378 & 45 & 66 & 0 & 1 & 975 & 0 & 1 & 21 & 3.7\% & 41.9\%\\ \hline
0.015$Z_{\odot}$ & 62 & 33 & 81 & 0 & 14 & 1100 & 0 & 14 & 36 & 3.2\% & 38.9\%\\ \hline
0.02$Z_{\odot}$ & 26 & 50 & 77 & 0 & 25 & 1136 & 0 & 25 & 55 & 3.2\% & 38.3\%\\ \hline
0.025$Z_{\odot}$ & 15 & 49 & 91 & 0 & 37 & 1168 & 0 & 34 & 74 & 3.2\% & 37.1\%\\ \hline
0.03$Z_{\odot}$ & 13 & 41 & 74 & 0 & 34 & 1227 & 0 & 29 & 71 & 3.4\% & 38.7\%\\ \hline
0.035$Z_{\odot}$ & 16 & 26 & 65 & 0 & 38 & 1241 & 0 & 21 & 57 & 3.6\% & 39.7\%\\ \hline
0.04$Z_{\odot}$ & 18 & 46 & 43 & 0 & 52 & 1281 & 0 & 34 & 39 & 3.8\% & 45.8\%\\ \hline
0.045$Z_{\odot}$ & 16 & 36 & 37 & 0 & 53 & 1164 & 0 & 21 & 28 & 3.7\% & 44.2\%\\ \hline
0.05$Z_{\odot}$ & 15 & 22 & 38 & 0 & 32 & 1205 & 0 & 20 & 34 & 3.7\% & 47.2\%\\ \hline
0.075$Z_{\odot}$ & 113 & 107 & 16 & 0 & 125 & 1013 & 0 & 88 & 15 & 3.8\% & 60.4\%\\ \hline
0.1$Z_{\odot}$ & 125 & 133 & 25 & 0 & 149 & 820 & 0 & 102 & 25 & 3.5\% & 64.0\%\\ \hline
0.125$Z_{\odot}$ & 169 & 129 & 11 & 0 & 150 & 689 & 0 & 101 & 11 & 3.4\% & 67.9\%\\ \hline
0.15$Z_{\odot}$ & 238 & 110 & 46 & 0 & 144 & 746 & 0 & 93 & 46 & 4.1\% & 76.5\%\\ \hline
0.175$Z_{\odot}$ & 148 & 84 & 68 & 0 & 101 & 826 & 0 & 73 & 68 & 4.1\% & 78.8\%\\ \hline
0.2$Z_{\odot}$ & 21 & 65 & 39 & 0 & 118 & 843 & 0 & 61 & 39 & 3.9\% & 80.3\%\\ \hline
0.225$Z_{\odot}$ & 31 & 54 & 27 & 0 & 126 & 627 & 0 & 52 & 27 & 3.1\% & 80.9\%\\ \hline
0.25$Z_{\odot}$ & 80 & 55 & 14 & 0 & 113 & 289 & 0 & 55 & 14 & 1.9\% & 67.9\%\\ \hline
0.275$Z_{\odot}$ & 90 & 61 & 14 & 0 & 119 & 194 & 0 & 60 & 14 & 1.6\% & 66.6\%\\ \hline
0.3$Z_{\odot}$ & 92 & 43 & 16 & 0 & 108 & 159 & 0 & 41 & 16 & 1.5\% & 68.5\%\\ \hline
0.325$Z_{\odot}$ & 59 & 62 & 3 & 0 & 143 & 140 & 0 & 62 & 3 & 1.5\% & 70.6\%\\ \hline
0.35$Z_{\odot}$ & 63 & 66 & 4 & 0 & 140 & 116 & 0 & 62 & 4 & 1.5\% & 77.0\%\\ \hline
0.375$Z_{\odot}$ & 63 & 60 & 2 & 0 & 137 & 67 & 0 & 57 & 2 & 1.3\% & 73.1\%\\ \hline
0.4$Z_{\odot}$ & 78 & 52 & 1 & 0 & 114 & 57 & 0 & 44 & 1 & 1.3\% & 79.8\%\\ \hline
0.425$Z_{\odot}$ & 105 & 45 & 0 & 1 & 79 & 46 & 1 & 31 & 0 & 1.2\% & 80.0\%\\ \hline
0.45$Z_{\odot}$ & 187 & 32 & 0 & 0 & 59 & 23 & 0 & 27 & 0 & 1.4\% & 83.7\%\\ \hline
0.475$Z_{\odot}$ & 177 & 38 & 1 & 0 & 57 & 22 & 0 & 33 & 1 & 1.4\% & 83.1\%\\ \hline
0.5$Z_{\odot}$ & 185 & 28 & 0 & 0 & 33 & 10 & 0 & 21 & 0 & 1.2\% & 80.7\%\\ \hline
0.75$Z_{\odot}$ & 1169 & 5 & 0 & 0 & 8 & 20 & 0 & 5 & 0 & 5.1\% & 98.7\%\\ \hline
$Z_{\odot}$ & 1034 & 2 & 0 & 0 & 5 & 18 & 0 & 2 & 0 & 3.5\% & 99.1\%\\ \hline
1.25$Z_{\odot}$ & 589 & 0 & 0 & 16 & 4 & 20 & 16 & 0 & 0 & 2.6\% & 98.2\%\\ \hline
1.5$Z_{\odot}$ & 925 & 0 & 0 & 25 & 3 & 20 & 25 & 0 & 0 & 4.1\% & 98.2\%\\ \hline
\end{tabular}}
\end{center}
\label{Table-ULXDCO}
\end{table*}

\end{document}